\documentclass[traditabstract]{aa}

\usepackage{graphicx}
\usepackage{xcolor}
\usepackage{multirow}

\usepackage{txfonts}

\usepackage{enumitem}

\begin{document}

   \title{Scaling relations at the central regions of nearby galaxies}

   \author{Patr\'icia da Silva
          \inst{1}
          \and
          R. B. Menezes\inst{2}
          \and
           T. V. Ricci\inst{3}
           \and
          F. Combes\inst{4,5}
           \and
          F. Pinna\inst{6,7}
          \and
          B. Barbuy\inst{1} 
          }

   \institute{ Instituto de Astronomia, Geof\'isica e Ci\^encias Atmosf\'ericas, Departamento de Astronomia, Universidade de S\~ao Paulo, Rua do Mat\~ao 1226, 05508-090, Butant\~a, SP,  Brazil\\
             \email{patricia2.silva@alumni.usp.br} 
         \and
            Instituto Mau\'a de Tecnologia, Pra\c{c}a Mau\'a 1, 09580-900, S\~ao Caetano do Sul, SP, Brazil
            \and 
            Universidade Federal da Fronteira Sul, Rua Jacob Reinaldo Haupenthal 1580, 97900-000, São Pedro, RS, Brazil
            \and
            Observatoire de Paris, LERMA, PSL University, CNRS, Sorbonne University, Paris, France
            \and
            Coll\`ege de France, 11 Pl. Marcelin Berthelot, 75231 Paris, France
            \and
            Instituto de Astrof\'isica de Canarias, C. V\'ia L\'actea, s/n, 38205 La Laguna, Santa Cruz de Tenerife, Spain
            \and
            Universidad de La Laguna, Pabellón de Gobierno, C/ Padre Herrera s/n, 38200, San Cristóbal de La Laguna, Santa Cruz de Tenerife, Spain
             }

   \date{Received 12 March 2025; accepted 11 August 2025.}

 
  \abstract{
  
Scaling relations between galactic parameters (e.g., luminosity, mass, metallicity, etc.) represent key pieces of evidence for investigating the processes of galaxy formation and evolution. In most studies, these relations have been obtained for large portions of the galaxies (i.e., on kiloparsec scales), but it is also important to evaluate these relations in smaller scales. In this work, we used optical data cubes of a subsample of nearby galaxies of the DIVING$^{3D}$ survey. These allowed us to analyze the scaling relations involving stellar velocity dispersion, stellar population age, and stellar population metallicity at the nuclear and circumnuclear regions of galaxies (within scales from tens to a few hundreds of parsecs). We detected correlations between the stellar velocity dispersion and the age, metallicity, and total stellar mass. These correlations are independent of galaxy inclinations, considering all morphological types, nuclear activity, and the presence or absence of galactic bars. We detected, for the first time, a correlation between the stellar velocity dispersion and stellar metallicity in the nuclear regions of galaxies. It is found to be qualitatively consistent with the well-known stellar mass-metallicity relation, described in previous studies, on kiloparsec scales. We also noted that barred galaxies tend to show younger and less metal-rich stellar populations than unbarred galaxies in the central regions, which may be a consequence of the bar triggering star formation in the nuclear regions of these objects. However, further studies, with larger samples, are necessary for comparisons between barred and unbarred galaxies, with the same mass and morphological types. Some active galactic nuclei (AGNs) in our sample are positioned above the observed correlation between stellar velocity dispersion and stellar population age, suggesting that their nuclear stellar populations are younger than expected. This may be a consequence of positive AGN feedback, triggering star formation. Conversely, starburst galaxies do not show nuclear stellar populations at ages over  one billion years.}


   \maketitle
%

\section{Introduction}\label{sec1}

Galaxies exhibit similarities in many of their physical properties, despite differences in mass, morphology, color, and other characteristics. A remarkable piece of evidence for these similarities is the set of observed scaling relations between various global parameters, such as luminosity, central supermassive black hole (SMBH) mass, stellar velocity dispersion, stellar mass, metallicity, and so on. These scaling relations offer valuable insights into the physical processes that influence galaxy formation and evolution.

Galaxy scaling relations have  been examined in many previous works (e.g., \citealt{fab76,tul77,cou07,kor13,sto21,aro23}). Measurements of the slope, zero point, and scatter of scaling relations (e.g., \citealt{cou07,kor13,lan15}) provide essential information about the structure of galaxies and also constraints for galaxy formation models (e.g., \citealt{dut17,sta19,van19}). It is worth mentioning, however, that most of these studies do not involve spatially resolved data. Analyses of scaling relations based on spatially resolved data could be very useful for understanding the physical
processes involved in the formation of galaxy structures; however, they are still lacking .

The analysis of scaling relations usually requires high-quality observations and also large samples of galaxies, to obtain statistically significant results. In principle, to avoid biases and selection effects,  data ought to be taken from a single survey, such as Sloan Digital Sky Survey (SDSS; \citealt{kol19}), Calar Alto Legacy Integral Field Area (CALIFA; \citealt{san12}), Spectroscopic Areal Unit for Research on Optical Nebulae (SAURON; \citealt{bac01}), ATLAS$^{3D}$ \citep{cap11}, and Sydney-AAO Multi-object Integral-field spectrograph (SAMI; \citealt{bry15}), among others.

A well-known scaling relation obtained for star-forming galaxies is the so-called mass-metallicity relation, involving the galactic stellar mass and metallicity (see \citealt{mai19} for a detailed review). The correlation between these two parameters was obtained from analyses involving color-magnitude diagrams of local ellipticals \citep{mcc68,san72,mou83,buo85}. Since its discovery, numerous studies of the mass-metallicity relation, both for quiescent and star-forming galaxies, have been performed with different samples of objects and models (e.g., \citealt{gal05,zah17,lia18,zha18, bulichi23}). Differences in the mass-metallicity relation detected in quiescent and star-forming galaxies have been interpreted as the result of quenching by ``starvation,'' which is the suppression of gas accretion (due to a number of physical processes), and the  primary mechanism to explain the stellar metallicity properties of galaxies \citep{pen15}.

Some studies have reported that this relation is qualitatively similar to the gas-phase mass-metallicity relation, which was first obtained, with a sample of nearby galaxies, by \citet{leq79}. Similarly to the stellar mass-metallicity relation, the gas-phase mass-metallicity relation has been exhaustively analyzed in a large number of works (e.g., \citealt{man10,lia15,cur20}). These studies helped determine that the mass-metallicity relation evolves with redshift (e.g., \citealt{mai08,yua13,zah14}), which allows the determination of the history of chemical enrichment in galaxies. Different mechanisms have already been proposed (listed by \citealt{mai19}) to explain the mass-metallicity relation. We briefly describe them below.

\begin{enumerate}[label=(\roman*),nosep]
    \item Galactic outflows that are globally related to stellar feedback, but located in the center of active galaxies dominated by active galactic nucleus (AGN) feedback. This makes them  more efficient in terms of removing metal-enriched gas from lower-mass galaxies than from higher-mass galaxies, due to the deeper gravitational potential well of the latter;
    \item Different evolutionary stages, with more massive galaxies evolving faster and rapidly increasing the metallicity of the interstellar medium at higher redshifts than lower-mass galaxies (so-called downsizing; \citealt{cow96,gav96,som15});
    \item Infall of metal-poor gas in lower mass galaxies coming from the intergalactic medium (IGM), which reduces their metallicity and induces star formation;
    \item Higher mass cut-off of the initial mass function (IMF) of stars in massive galaxies, which results in the observed mass-metallicity relation;
    \item Higher metallicity of the accreted gas (caused by previous star formation episodes and outflows) in more massive galaxies, as compared to lower-mass galaxies.  \\
\end{enumerate}

In this work, we study scaling relations involving stellar velocity dispersion, stellar population age, stellar population metallicity, and galactic stellar mass, in the central regions of galaxies. We used data cubes obtained with the Deep Integral Field Spectroscopy View of Nuclei of Galaxies  (DIVING$^{3D}$ -- \citealt{diving3d}) survey. For the purposes of this work, we used a subsample of the DIVING$^{3D}$ survey, called mini-DIVING$^{3D}$ (see \citealt{men22}; hereafter, Paper I). This sample comprises all galaxies of the  DIVING$^{3D}$ brighter than $ B=  11.2$. The spatial scale considered in this work is about 300 pc (circumnuclear) around the nuclear region (tens of pc).

This paper is structured as follows. In Section \ref{sec2}, we describe the methods and data used in this work. In Section \ref{sec3}, we present the results, which are discussed in Section \ref{sec4}. Section \ref{sec5} summarizes  the conclusions of our analysis.

\section{Methodology}\label{sec2}

The DIVING$^{3D}$ survey is aimed at observing the nuclear regions of all galaxies in the southern hemisphere brighter than $ B=  12.0$ and with Galactic latitudes of $|b| > 15^{\circ}$, with the integral field unit (IFU)\ of the Gemini Multi-Object Spectrograph (GMOS) at the Gemini-North and Gemini-South telescopes. The resulting data enable us to analyze the nuclei of galaxies, at a high spatial and spectral resolution (for more information, see \citealt{diving3d}),  consisting of data cubes in the optical band (for late type galaxies: 4700 \AA\ to 6900 \AA\ with spectral resolution of 1.2 \AA; for early type galaxies: 4030 \AA\ to 7100 \AA\ with spectral resolution of 1.8 \AA), with spatial scales from about 50 to several hundreds of parsecs and spatial resolution of about 0.7 arcsec. However, it is worth mentioning that due to certain instrumental issues, some data cubes presented ``artifacts'' at the start or at the end of the wavelength coverage. These compromised portions of the data cubes were removed, which resulted in slightly shorter wavelength coverages for these objects; however, this did not affect the range of the main emission lines of the optical bands.

From  this survey, we also obtained optical data cubes from the SOAR Integral Field Spectrograph (SIFS), at the SOAR telescope to complete the sample. In this case, the spectral resolution was 1.3 \AA\ and the spatial resolution was about 1 arcsec (with a FoV likely closer to 1 kiloparsec). 

In this work, we use the sample of galaxies brighter than $ B=  11.2$, which we call mini-DIVING$^{3D}$ (Paper I). Table \ref{table1} shows the sample of 57 galaxies, with different morphological types and distances, their properties, and spatial scales of the field of view (FoV). Within the mini-DIVING$^{3D}$ sample, with exception of NGC 1232 (obtained with SIFS), all galaxies were observed with Gemini North and South telescopes. Their data are available in the Gemini public data archive. 

The observation details, data reduction, and treatment are described in Paper I, following the method developed by \citet{men14,men15,menezes19}. In the case of GMOS data cubes, the data were reduced using the Gemini package, in Image Reduction and Analysis Facility (\textsc{iraf}) environment, developed by the Gemini team, which includes: trim determination, bias subtraction, cosmic ray removal \citep{van01}, correction for gain variations, flux and wavelength calibrations, and construction of data cubes with spatial pixels (spaxels) of 0.05 arcsec. The treatment includes correction of differential atmospheric refraction, Butterworth spatial filtering \citep{gwoods}, instrumental fingerprint removal, and Richardson-Lucy deconvolution \citep{rich,lucy}. For the SIFS data, similar reduction process and the same procedures were applied (see Paper I), resulting in data cubes with spaxels of 0.1 arcsec, with the exception of the differential atmospheric refraction correction, since this instrument has an atmospheric dispersion corrector (ADC).

\subsection{Nuclear and circumnuclear spectra}

In this work, we show the analysis performed on two spectra for each galaxy: the nuclear and the circumnuclear spectra. The nuclear spectrum of each galaxy of the sample was extracted from a circular region, centered at the stellar nucleus of the object (defined by the image of the stellar continuum of the data cube), with a radius equal to half of the full width at half maximum (FWHM) of the point spread function (PSF) of the treated data cube. The radii of the extraction regions (converted to pc, taking into account the distances of the galaxies) are shown in Table~\ref{table1}. The FWHM values of the PSFs of all observations are shown in Table 1 of Paper I. We also multiplied the extracted spectrum by a constant factor, in order to correct the flux values for the imprecision introduced by the fact that the spectrum was extracted from a truncated circular region, although the PSF has a more extended  Gaussian shape. For further details on this flux correction and how the PSFs of the data cubes were estimated, we refer to Paper I. The circumnuclear spectrum of each object was taken as the total spectrum of the corresponding data cube minus the nuclear extracted spectrum. The typical stellar continuum signal-to-noise ratio (S/N) is $\sim$ 30 for the extracted nuclear spectra and $\sim$ 10 for the extracted circumnuclear spectra, generating higher uncertainties in the circumnuclear results when compared with the nuclear regions.

A natural question at this point is related to the fact the PSF is not a physical quantity and corresponds to different physical scales, depending on the distance of the galaxy. Thus, we must consider what  impact the use of this parameter has on our ability to distinguish between nuclear and circumnuclear regions. We start with the assumption that any possible effects of this approach on the results of our work are likely to be secondary. For the nearest galaxies, for example, there is the possibility that the extracted circumnuclear spectra include emission from the nuclear region. However, the area corresponding to the circumnuclear region is considerably larger than the one the contaminating nuclear emission was extracted from. As a consequence, in these cases, the contamination of the extracted circumnuclear spectra by nuclear emission is probably secondary. On the other hand, for the farthest galaxies in our sample, there is the possibility that the extracted nuclear spectra include circumnuclear emission. However, since the flux of the spectra from the actual nuclear region is considerably higher than that of the circumnuclear region, the contamination of the nuclear spectra by circumnuclear emission is also likely to be secondary.

\subsection{Spectral synthesis}\label{sec_spectral_synt}

Before any spectral synthesis was applied, the extracted spectra were corrected for  Galactic interstellar extinction, using the $A_V$ values from \citet{sch11} and the extinction law of \citet{car89}, then passed to the rest frame, using the redshift values from NASA Extragalactic Database (NED\footnote{The NASA/IPAC Extragalactic Database (NED) is funded by the National Aeronautics and Space Administration and operated by the California Institute of Technology} - and references therein). The spectra were also re-sampled to spectral pixels of 1 \AA, to match the base spectra used in the spectral synthesis.

To obtain the stellar parameters (e.g., age, metallicity, and stellar velocity dispersion), we performed a spectral synthesis, which consists of fitting the observed stellar spectra with combinations of template stellar population spectra from a base. The base we employed combines the Granada models of \citet{gon05} for ages lower than 63 Myr and the models of \citet{vaz10} for ages that are even higher. This base was obtained with the Salpeter Initial Mass Function and the evolutionary tracks of \citet{gir00}; except for ages lower than 3 Myr, which were obtained from Geneva tracks \citep{sch92,sch93a,sch93b,cha93}. The base contains 150 spectra of stellar populations, with ages between $1.0 \times 10^6$ and $1.3 \times 10^{10}$ yr and metallicities between 0.0001 and 0.05 (Z$_{\sun}$ = 0.02 being the solar metallicity). These values of metallicity (Z) actually represent mass fractions of metals (linear metallicities). We chose between two methods, depending on the respective parameters we wanted to determine:
\begin{enumerate}[label=(\roman*),nosep]
    \item To obtain the stellar velocity dispersion ($\sigma$) of each extracted spectrum, we applied the penalized pixel fitting (pPXF, \citealt{cap17}) procedure, since it gives us precise kinematic parameters.We took into account the spectral resolution of the stellar population base, which is $\sigma_{base}$ = 48.8 km~s$^{-1}$, along with the spectral resolution of the observed spectra, which is $\sigma_{inst}$ = 41.1 km~s$^{-1}$ for early type galaxies and $\sigma_{inst}$ = 29.7 km~s$^{-1}$ for late type galaxies. The $\sigma_{inst}$ values were estimated from the [OI]$\lambda$5577 emission line. No corrective polynomials were added when applying the pPXF technique. The uncertainties of $\sigma$ were obtained with a Monte Carlo procedure. First, for each spectrum, the pPXF was applied and the provided synthetic stellar spectrum (corresponding to the fit obtained by the method) was subtracted from the observed one, resulting in a spectrum containing only emission lines and spectral noise. Then, we constructed a histogram of the spectral noise, based on wavelength intervals without emission lines, and fitted a Gaussian function to it. With that, we generated different samples of random noise, following the same function fitted to the histogram of the spectral noise. These noise distributions were added to the original synthetic stellar spectrum provided by pPXF, which resulted in ``noisy spectra.'' This process was repeated 100 times. Finally, the pPXF was applied to each of these noisy spectra and the uncertainty of $\sigma$ was taken as the standard deviation of the $\sigma$ values obtained with all these applications of the pPXF technique.
    
    \item To derive the stellar population parameters from the extracted spectra, we applied the \textsc{starlight} software \citep{cid05}, since it is optimized to give age and metallicity of the stellar populations. The mean age and metallicity of each spectrum were calculated as weighted means, with the flux fraction of each stellar population taken as the weight. It is worth mentioning that this determination of the stellar population parameters did not require that the $\sigma$ values were kept constant. The reason is that we performed a few tests with this procedure, while keeping and without keeping $\sigma$ constant (and equal to the value provided by the pPXF technique). Finally, we verified that these two procedures resulted in compatible stellar population parameters. The uncertainties of the mean ages and metallicities were obtained with the same Monte Carlo procedure used in the case of $\sigma$.\\
\end{enumerate}

\begin{figure*}
\begin{center}

 \includegraphics[scale=0.3]{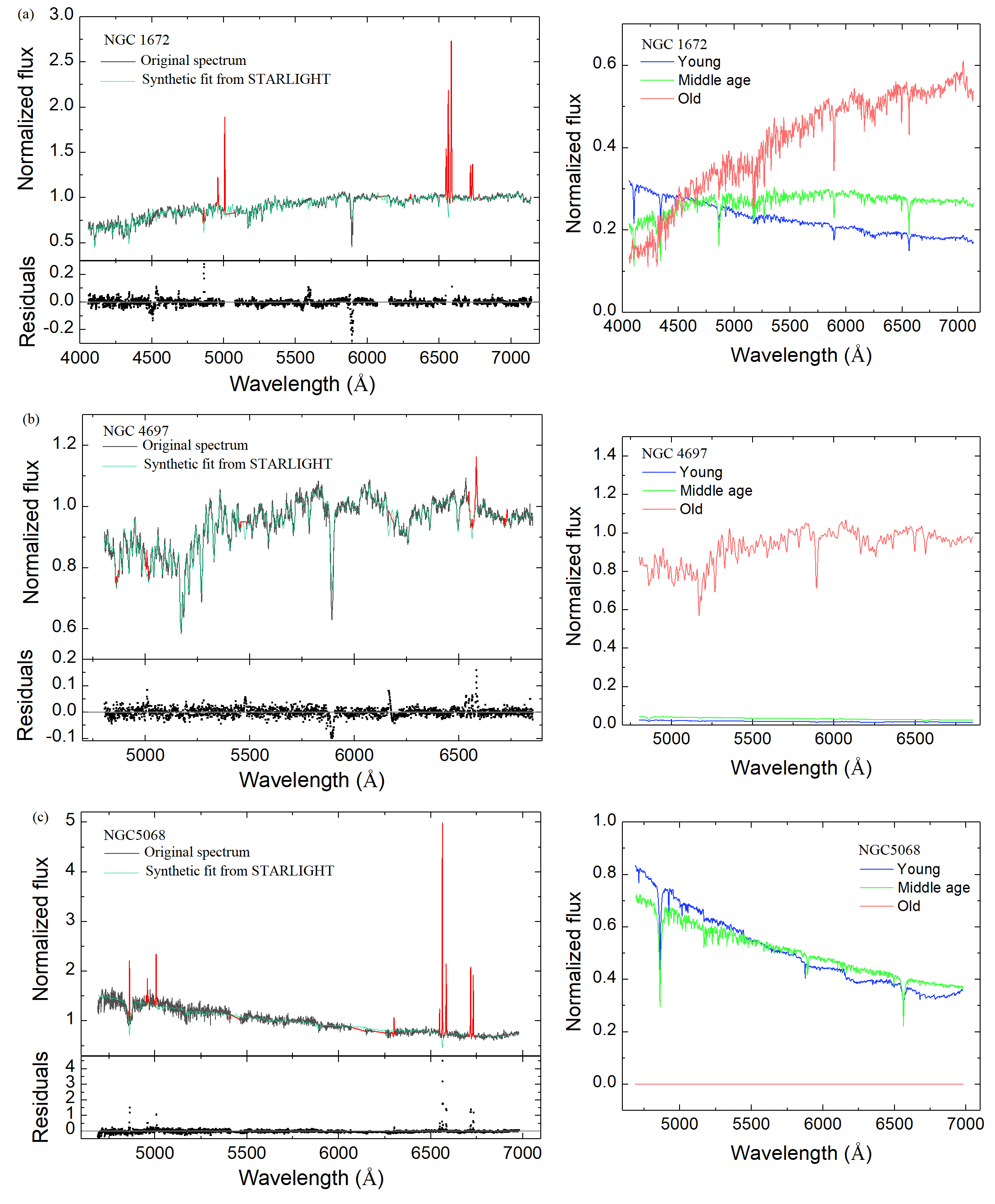}
\caption{  Left: Spectral synthesis fits from \textsc{starlight} and residuals of the nuclear spectra of 
three objects: (a) NGC 1672 (early spiral), (b) NGC 4697 (early type), and (c) NGC 5068 (late type). The fluxes are normalized according to the \textsc{starlight} normalization.  Dark green lines represent the synthetic total fit provided by the \textsc{starlight} software and red lines correspond to the masked areas, not considered in the fitting. Right: Stellar population components of the fits on the left. The red color represents old stellar populations ($\sim 2.5 \times 10^9 $ to $\sim 1.8 \times 10^{10}$ yrs ), green lines correspond to middle age stellar populations ($\sim 5.5 \times 10^7 $ to $\sim 1. 4\times 10^{9}$ yrs), and blue lines represent young stellar populations  ($ \sim 1 \times 10^6 $ to $\sim 4\times 10^{7}$ yrs) fits.\label{spectralsytnthesis}}
  
\end{center}
\end{figure*}

The results provided by spectral synthesis algorithms are usually affected by degeneracies. As a consequence, the values and the corresponding uncertainties obtained with the pPXF technique and the \textsc{starlight} software must be taken with caution. On the other hand, since the same procedures (with the same base) were applied to all the extracted spectra to derive $\sigma$ and the mean ages and metallicities of the stellar populations, we could conclude that any possible errors introduced by the methodologies affected all the spectra in the same way. Therefore, although the absolute values of the parameters determined for the different spectra may have been affected by these systematic errors, a comparison between such parameters can still be considered reliable for this work. In Fig.~\ref{spectralsytnthesis}, we show the resulting fits of both procedures for three objects of our sample. 

Since the spectra obtained for the early type and late type galaxies in our sample have different wavelength ranges (as mentioned in Section 2), a natural question at this point relates to whether the spectral synthesis (performed with the \textsc{starlight} software) applied to the spectrum of an early type galaxy in our sample that is cropped to have the same range of the spectrum of a late type galaxy would provide different results from those obtained with the original (uncropped) spectrum. To attempt to answer this question, we performed the following test: we cropped the spectra of the early type galaxies to match the wavelength range of the late type galaxies. Then, we applied the spectral synthesis, following the same procedure described above. The mean ages and metallicities we obtained were compatible, within the uncertainties, with those provided by the procedure applied to the spectra with the original wavelength ranges. On this basis, we were able to conclude that the different wavelength ranges of the spectra of early type and late type galaxies did not significantly affect the results presented in this work.

\subsection{Division of the sample}

The galaxy sample has many features worth exploring. We made divisions based on their main characteristics and the interpretation of their impacts on the stellar evolution of their center: nuclear activity, morphological type, and the presence of bars. Table \ref{table1} shows the nuclear activity classification as it is given in Paper I.  Some objects are not classified, some have more than one classification, and some are very ambiguous. In order to separate these objects in different categories regarding their nuclear activity, we took into account the information in Table 6 of Paper I, stating which ones have observational features typical of AGNs. In this work, low-ionization nuclear emission-line regions (L: LINERs) that have a point-like hard X-ray emission are classified as AGNs. LINERs without such an emission are classified as ``unconfirmed AGNs.'' Objects identified as HII/T/L (HII: starburst galaxy; T: transition object; L: LINER) in Table \ref{table1} and with a point-like hard X-ray emission are included in the category of AGNs, as well as the ones that have L/S (LINER, S: Seyfert) classifications. NGC 1232 has a point-like hard X-ray emission; however, the optical data were not at a high enough spatial resolution to enable a reliable analysis of its activity. In this case, we included it in the ``unconfirmed AGNs'' group. NGC 2442 is classified as LINER and, in Table 6 of Paper I, it was described as not having hard X-ray emission; however, this object was included in the ``AGNs'' category, as \citet{dasilva21} reported it as an AGN. An analogous case is  that of NGC 6744, where there is a strong evidence in \citet{dasilva18} that this galaxy harbors an AGN. NGC 157 is classified as a transition object (T, between starburst and active galaxies) in Paper I; however, a more detailed study in \citet{patricia20} showed that it has a starburst nucleus. The objects classified as HII/L or HII/L/T/S were included in the group labeled ``unclassified.'' The objects that do not have enough emission lines to allow a classification based on the BPT diagram \citep{bal81} were included in the group called ``no-emission.'' ``unconfirmed AGNs,'' transition objects, and ``unclassified'' represent each $\sim$ 5\% of the sample (three galaxies each). About 60\% of the galaxies in the sample are AGNs, totaling 34 galaxies. Starburst galaxies represent 18\% of the sample (ten galaxies), while  $\sim$ 7\% are ``no-emission'' galaxies (four galaxies). See Table \ref{tableproperties} for the classification adopted in this work for each galaxy of the sample. 

We also separated the sample regarding its morphological types: early types (ellipticals and S0: 19 galaxies, $\sim$ 33\%), early spirals (S0/a - Sbc: 21 galaxies, $\sim$ 37\%), late spirals (Sc - Sd: 17 galaxies, $\sim$ 30\%), and barred (30 galaxies, $\sim$ 53\%) and unbarred galaxies (27 galaxies, $\sim$ 47\%), according to the morphological types given in Table \ref{table1}. 

The results were separated according to the region of the extracted spectrum: nuclear and circumnuclear, taking into account the characteristics of the host galaxies. We also considered, according to the \citet{hoyer2021} catalog, the galaxies of our sample that have confirmed nuclear stellar clusters (NSC),  representing $\sim$ 17\% of the sample (nine galaxies: NGC 247, NGC 300, NGC 3115, NGC 3621, NGC 5068, NGC 5102, NGC 5236, NGC 7090, and NGC 7793).

In the next sections, we analyze plots of log($\sigma$) versus log(age) and log($\sigma$) versus log Z both for the nuclear and circumnuclear spectra of the objects in the sample. Here, $\sigma$ is the mean stellar velocity dispersion, ``age" is the weighted mean value of the stellar population ages, and Z is the weighted mean value of the stellar metallicity for each spectrum. It is worth emphasizing that (as explained in Section \ref{sec_spectral_synt}) the weighted mean values of age and Z were calculated using the flux fractions of the different stellar populations as weights.

To evaluate whether any possible correlations in the previous plots could be observed by replacing $\sigma$ by the total stellar mass of the galaxy ($M_*$, obtained from the literature; see Table \ref{table1} for references), we also analyzed the plots of log($\sigma$) versus log($M_*$). 

Finally, we also evaluated the inclinations of the galaxies and separated them in two groups: higher than 65 $^{\circ}$ (18 galaxies: $\sim$ 32\% of the sample) and lower than 65$^{\circ}$ (39 galaxies: $\sim$ 68\% of the sample) to see its influence on the parameters we sought to observe. The inclinations were taken from Hyperleda\footnote{http://leda.univ-lyon1.fr/} and are shown in Table \ref{table1}.

\section{Results}\label{sec3}

\subsection{Nuclear spectra}

\begin{figure*}
\begin{center}

 \includegraphics[scale=0.3]{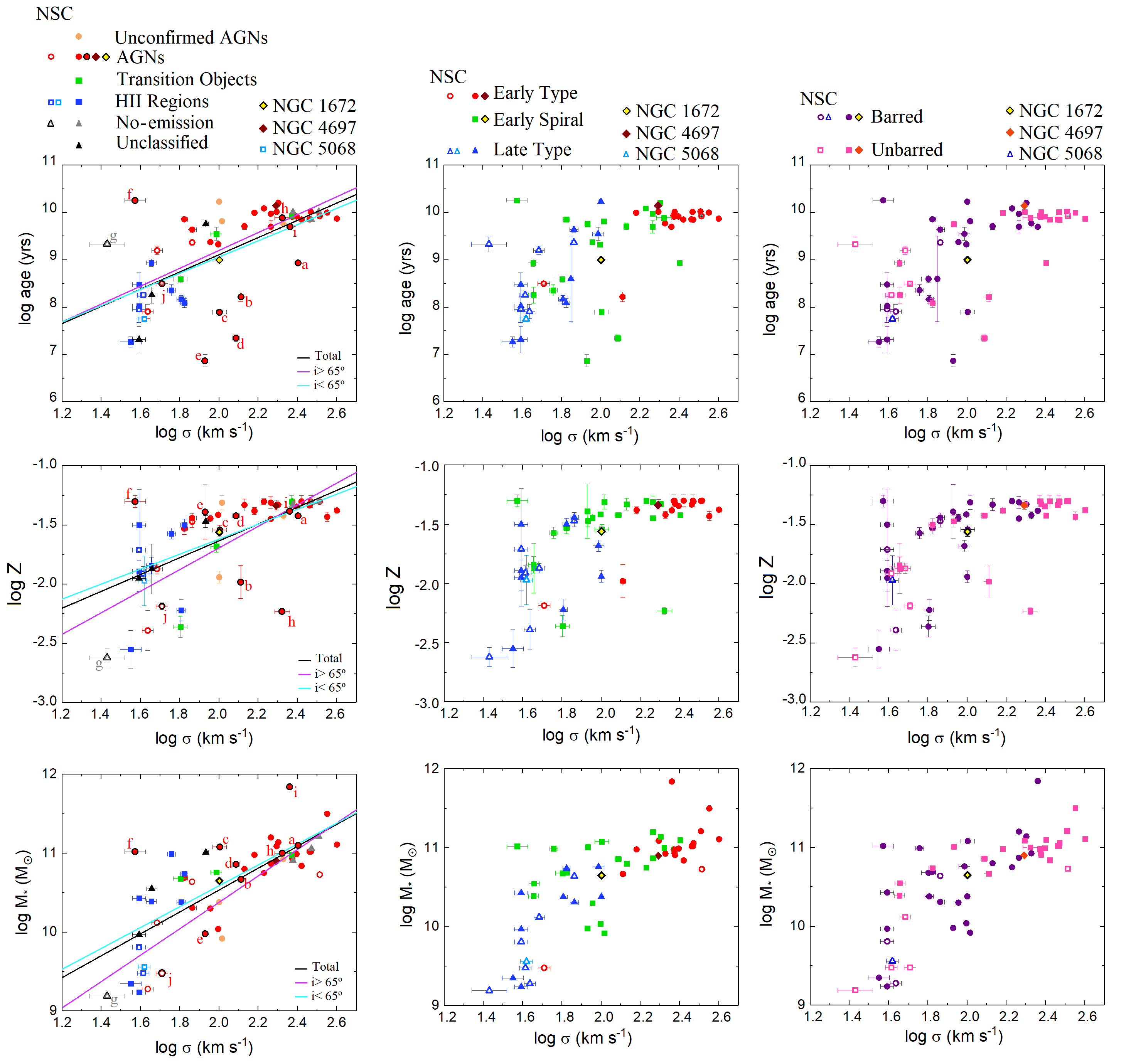}
\caption{Age, metallicity, and total stellar mass versus the nuclear stellar velocity dispersion. The sample is divided in groups, according to their features: nuclear activity, morphology, and bar presence, each one shown in the captions of each panel. According to \citet{hoyer2021}, this sample of galaxies contains nine galaxies with nuclear stellar clusters, which are represented by the open points. Panels on the left show the fits considering the total sample (black), sample with inclinations higher than  65$^{\circ}$ (purple), and sample with inclinations lower than  65$^{\circ}$ (cyan). There is also the indication of three galaxies used as example in Fig.~\ref{spectralsytnthesis}. Letters a, b, c, d, e, f, and g next to points represent the following outliers of the plot log (age) versus log ($\sigma$): NGC 4594 (a), NGC 5128 (b), NGC 613 (c), NGC 1068 (d), NGC 1566 (e), NGC 1365 (f), NGC 300 (g). Here, h corresponds to the outlier NGC 7213 in the plot of log(Z) versus log ($\sigma$),   i represents the outlier NGC 1316 in the plot of log($M_*$) versus log ($\sigma$), and  j corresponds to NGC 5102, which is an outlier in the plots of morphological types.}\label{sigma_plots_nuclear}
  
\end{center}
\end{figure*}

Figure~\ref{sigma_plots_nuclear} displays the plots of log(age) versus log($\sigma$), log(Z) versus log($\sigma$), and log($M_*$) versus log($\sigma$) for the nuclear spectra. In all plots, there appears to be a $\sigma$ threshold for active and non-active galaxies, near log($\sigma$) $\sim$ 1.8 ($\sim$ 65 km s$^{-1}$); meaning that higher nuclear $\sigma$ values than this  can indicate (at least) a higher probability of the presence of an AGN. The early type galaxies are all located above log($\sigma$) $\sim$ 2.1 ($\sim$ 130 km s$^{-1}$), with the exception of NGC 5102 (j point in the graphs). This might be due to the fact that this galaxy has an unusual star formation and strong evidence of being the result of a merger (e.g., \citealt{beaulieu10,davidge15,mit16}) or to inaccuracies related to the fitting of its spectra. For the early spiral galaxies, there is an upper limit for the nuclear $\sigma$ of log($\sigma$) $\sim$ 2.4 ($\sim$ 250 km~s$^{-1}$). We also see that the barred galaxies have mostly lower nuclear $\sigma$ (below log($\sigma$) = 2.4, $\sim$ 250 km~s$^{-1}$). Finally, all starbursts are in the category of late type galaxies, with exception of NGC 157, which is an early spiral.

The plots of log(age) versus log($\sigma$) in the top panels of Fig.~\ref{sigma_plots_nuclear} show a positive correlation between these two parameters. The most massive galaxies are the ones that have the highest $\sigma$ and the oldest stellar populations. We do not see galaxies with nuclear emission typical of HII regions with stellar population ages older than $\sim$ one billion years. Galaxies with no emission are in the oldest part of the diagram as well, which may indicate that these galaxies have low gas content, leading to no emission and no considerable star formation activity. We can see that there are also active galaxies that have young stellar populations (younger than one billion years), representing 32\% of the AGNs sample. Meanwhile, the late spiral galaxies show a wide range of ages and the early type galaxies are composed of old stellar populations, as expected, with ages higher than 3.16 billion years (log(age) $\sim$ 9.5 yrs), in general. The mean ages of 47\% of the barred galaxies are lower than 1 Gyr. On the other hand, 70\% of the unbarred galaxies show mean ages higher than 1 Gyr.

The plots of log(Z) versus log($\sigma$) in the middle panels of Fig.~\ref{sigma_plots_nuclear} show (again) a positive correlation between these two parameters. Galaxies with AGNs have higher metallicities (higher than log(Z) = -1.6, mostly around log(Z) $\sim$ -1.3). We also see non-active nuclei with a wide range of metallicities. The same seems to happen to the transition objects of the sample. Early type galaxies have the highest metallicities of the sample and late spirals have a wide range of metallicities. However, the late spirals are the ones that have the lowest values of metallicity of the sample. NGC 5102 (point j) and NGC 5128 (point b) are the only two early spirals with low values of metallicity, possibly due to the merger events, which they both show evidence of \citep{beaulieu10,davidge15,abdollahi24}. 
Regarding the barred and unbarred classifications, we cannot see any clear difference.

The plots of log($M_*$) versus log($\sigma$) in the bottom panels of Fig.~\ref{sigma_plots_nuclear} also show a positive correlation between both quantities. When we look at the morphological types of the galaxies in the sample, we see that (as expected) early type galaxies fall in the most massive and, consequently, the higher $\sigma$ part of the diagram, while the late spirals fall in the opposite side. The late spiral galaxies seem to have an upper limit for the total stellar mass, log($M_*$) = 10.8 $M_{\bigodot}$, and an upper limit for $\sigma$, $\sim$ 100 km s$^{-1}$. Regarding the presence of the bar, we can see that the unbarred galaxies are the most massive ones, in general.

\begin{figure}
\begin{center}

 \includegraphics[scale=0.35]{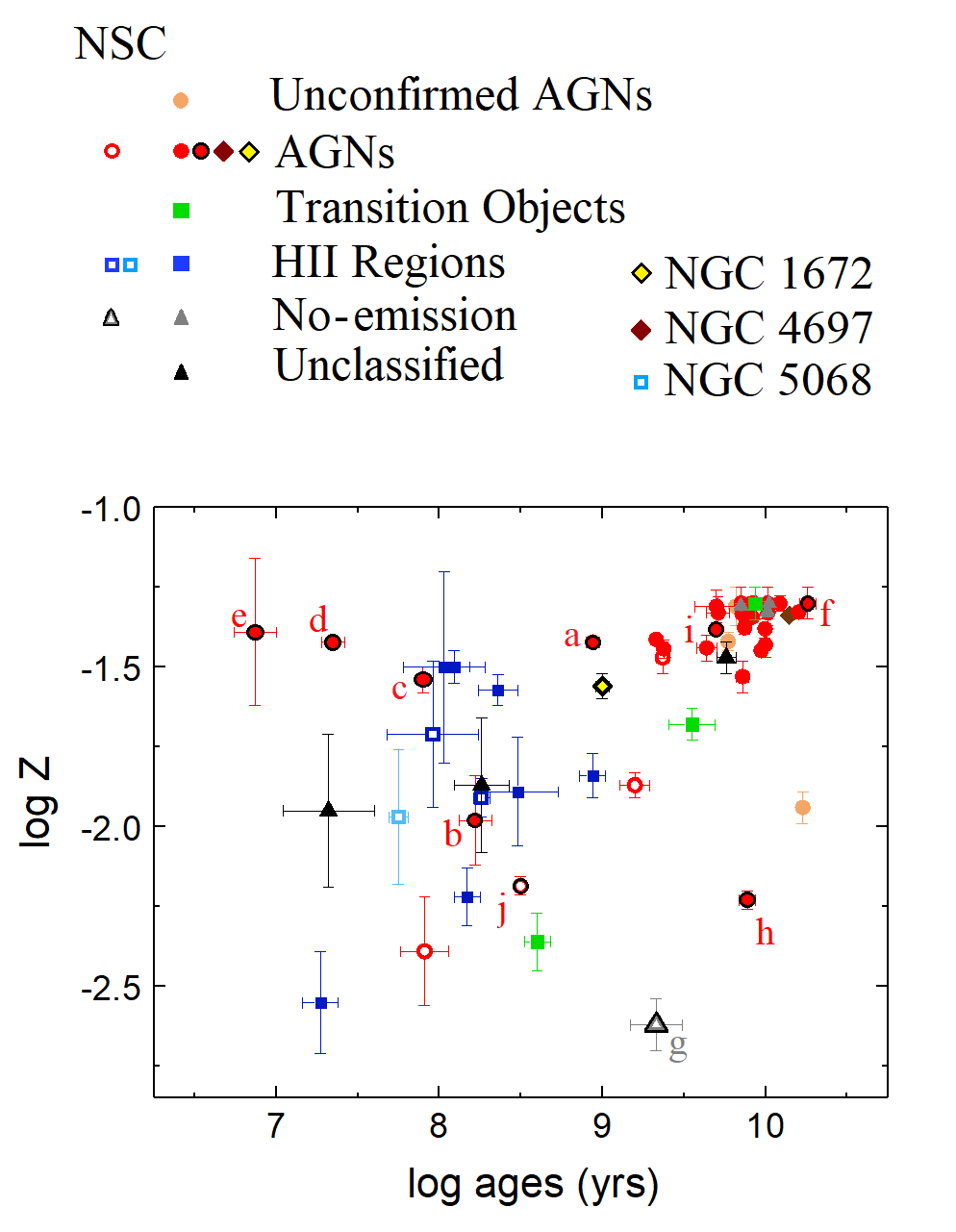}
\caption{ Metallicity versus age of the nuclear spectra of the galaxies of the sample, considering the BPT classification. Letters a to j represent the same outliers of Fig.~\ref{sigma_plots_nuclear}:NGC 4594 (a), NGC 5128 (b), NGC 613 (c), NGC 1068 (d), NGC 1566 (e), NGC 1365 (f), NGC 300 (g), NGC 7213 (h), NGC 1316 (i), and NGC 5102 (j). \label{metal_age_nuc}}
  
\end{center}
\end{figure}

Considering the positive correlations observed in the graphs in Fig.~\ref{sigma_plots_nuclear}, we evaluated the possibility of fitting functions to these graphs. Polynomial functions were the ones that provided the best fits; however, due the limited number of galaxies in our sample, we could not ensure that certain features in the plots were real. Taking this into consideration and to avoid any overfitting, we opted to perform simple linear fits to the graphs, which are shown in Fig.~\ref{sigma_plots_nuclear}. For completeness, third degree polynomial fits to these graphs (which can be taken as an upper limit for the order of polynomials to be fitted to the points) are shown in Appendix \ref{app_3d}. At first, we tried to apply these fits taking into account the uncertainties of the individual measurements, which is equivalent to give more weight to the points with lower uncertainties. However, we noted that the resulting fits did not pass through most of the points in the graphs (in other words, the fits did not seem to be consistent with the observed distribution of points). A possible explanation is that some of the uncertainties of the points may be overestimated. Considering this, we opted to perform the fits without such uncertainties and, as a consequence, the uncertainties of the parameters of the fitted functions only take into account the scattering of the points. We can see that the linear functions reproduce (with a good level of precision), the positive correlations between the parameters.  To evaluate any possible influence of the inclination of the galaxies, relative to the line of sight, on the correlations detected in Fig.~\ref{sigma_plots_nuclear}, we performed two additional linear fits: one taking into account galaxies with inclinations higher than 65\degr and the other considering only galaxies with inclinations lower than 65\degr. The parameters of all linear fits are shown in Table~\ref{coef_nuclear}. For all graphs, the angular and the linear coefficients obtained for all galaxies, for galaxies with inclinations higher than 65\degr, and for galaxies with inclinations lower than 65\degr~are compatible, at the 1-$\sigma$, 2-$\sigma,$ or 3-$\sigma$ levels.

\begin{table*}
\centering
\caption{Coefficients of the linear fits in Fig.\ref{sigma_plots_nuclear}.}
{%
\begin{tabular}{ccccccc}
\cline{2-7}
 & \multicolumn{2}{c}{ log(age)$\times$ log$\sigma$} & \multicolumn{2}{c}{ Z $\times$ log$\sigma$} & \multicolumn{2}{c}{ log(M$_*$) $\times$ log$\sigma$} \\ \cline{2-7} 
Sample & \begin{tabular}[c]{@{}c@{}}Angular \\ Coefficient\end{tabular} & \begin{tabular}[c]{@{}c@{}}Linear\\ Coefficient\end{tabular} & \begin{tabular}[c]{@{}c@{}}Angular \\ Coefficient\end{tabular} & \begin{tabular}[c]{@{}c@{}}Linear\\ Coefficient\end{tabular} & \begin{tabular}[c]{@{}c@{}}Angular \\ Coefficient\end{tabular} & \begin{tabular}[c]{@{}c@{}}Linear\\ Coefficient\end{tabular} \\ \hline
Total                  & 5.5$\pm$ 0.6                                               & 1.8 $\pm$ 0.3                                               & -3.05 $\pm$ 0.23                                               & 0.71 $\pm$ 0.11                                              & 7.8 $\pm$ 0.3                                               & 1.38 $\pm$ 0.16                                               \\
i > 65$^{\circ}$ & 5.4 $\pm$ 0.6                                             & 1.9 $\pm$ 0.3                                               & -3.5 $\pm$ 0.4                                               & 0.91 $\pm$ 0.16                                                & 7.0 $\pm$ 0.6                                               & 1.7 $\pm$ 0.3                                               \\
i < 65$^{\circ}$    & 5.6 $\pm$ 0.9                                               & 1.7 $\pm$ 0.4                                                & -2.8 $\pm$ 0.3                                               & 0.6 $\pm$ 0.15                                              & 8.0 $\pm$ 0.4                                                 & 1.32 $\pm$ 0.20                                              \\ \hline
\end{tabular}
}

\label{coef_nuclear}
\end{table*}

Figure~\ref{metal_age_nuc} shows the relation between age and metallicity in the nuclear regions of the objects in the sample and the classification of their nuclear emission. We can clearly see that most of the galaxies classified as AGNs, including the unconfirmed AGNs category, fall in the same area of the plot: with higher metallicities and ages. However, there are some exceptions. Apart from NGC 5236, all AGNs with nuclear star clusters (see further detail in Section 4) are far from this area, together with the outliers we identified in the plots of Fig.~ \ref{sigma_plots_nuclear}. This suggests the occurrence of specific starbursts in these objects that make their nuclear regions younger and sometimes, depending on how these nuclear regions are being fed, with lower metallicities. An interesting point is that the starburst galaxies have an upper limit for weighted mean ages (one billion years) and we can find AGNs with nuclear stellar populations ages lower than one billion years (e.g., NGC 1566:\ point e in the graph). However, we cannot find starburst galaxies with ages that are higher than one billion years. Thus, it is very likely that the objects whose nuclear emission classification as an AGN is unclear are not starburst nuclei, since they are older than one billion years: all unconfirmed AGNs (that also have high metallicities), some transition objects, all no-emission galaxies (which could be due to very high obscured nuclei or recent turned-off AGNs), and one unclassified galaxy (NGC 4030).

\begin{figure*}
\begin{center}

  \includegraphics[scale=0.32]{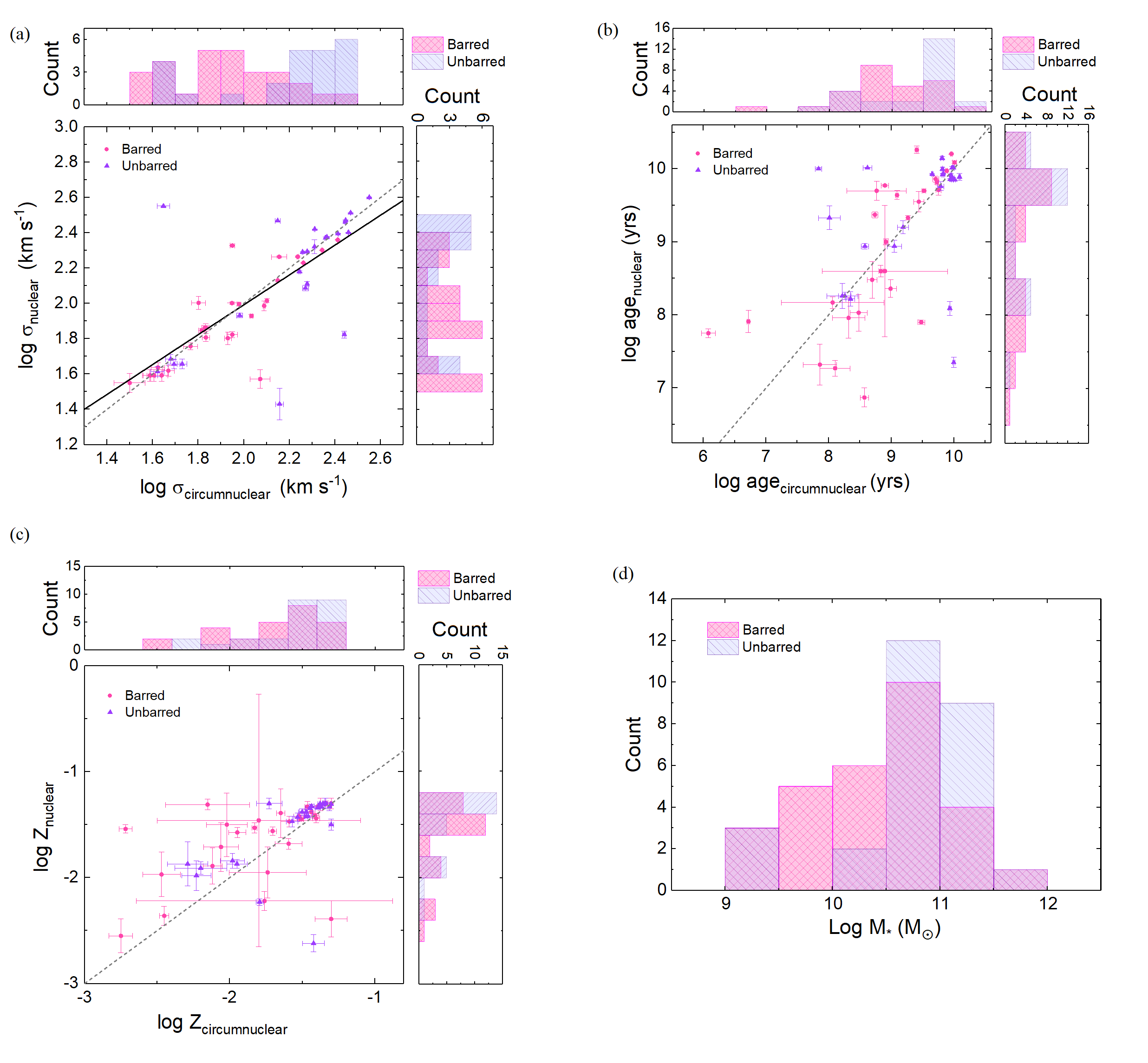}
  \caption{Plots of nuclear versus circumnuclear and histograms of (a) stellar velocity dispersion, (b) age, and (c) metallicity of the sample separated in barred and unbarred galaxies. The histogram of the mass of the barred and unbarred galaxies is shown in panel (d). The black line corresponds to the linear fit, which was only made for the stellar velocity dispersion, since its distribution is close to linear. The dashed gray line is y=x, for reference. } \label{histograms}
  
\end{center}
\end{figure*}

The histograms in Fig.~\ref{histograms} give us the main properties of the sample divided in barred and unbarred galaxies. 
Regarding nuclear $\sigma$ (Fig.~\ref{histograms}a), the barred galaxies show a threshold around 240 km s$^{-1}$ and peak in lower values, while the unbarred galaxies have a peak around 250 km s$^{-1}$ and reach higher values. The interesting fact about this plot is that we can clearly see the correspondence between the values of stellar velocity dispersion in the nuclear and circumnuclear regions ($log \sigma_{nuclear}= (0.85 \pm 0.10)*log \sigma_{circumnuclear} +(0.30 \pm 0.20)$). For the nuclear stellar population ages (Fig.~\ref{histograms}b), the peaks of the histograms for barred and unbarred galaxies are close to log(age) = 9.75 yrs; however, only the distribution in barred galaxies includes younger ages. In the circumnuclear region, the peaks differ a little bit: the barred galaxies have a peak near log(age) = 8.75 yrs and the unbarred galaxies still near log(age) = 9.75 yrs, which shows the consistency of the properties of circumnuclear regions in unbarred galaxies and the star formation more present in barred galaxies. Regarding the stellar metallicity (Fig.~\ref{histograms}c) in the nuclear region, we can see that it is generally lower in barred galaxies than in unbarred galaxies. There is not much difference regarding the circumnuclear values. Finally, the unbarred galaxies show peaks between log($M_*$) = 10.5 $M_{\bigodot}$ and  log($M_*$) = 11.5 $M_{\bigodot}$, while the barred galaxies have a peak around log($M_*$) = 10.7 $M_{\bigodot}$, with a higher concentration towards lower masses (Fig.~\ref{histograms}d). 

The histograms in Fig.~\ref{histograms} show that in the circumnuclear region, barred galaxies tend to present lower values of $\sigma$ than unbarred galaxies, which is the same trend observed in the nuclear region histogram. Regarding the circumnuclear stellar population ages, there is a clear peak of about 10 gigayears (Gyrs) for the unbarred galaxies, the same trend observed in the nuclear regions. For barred galaxies, the peak is close to 10 Gyrs in the nuclear region and close to 1 Gyr in the circumnuclear region. That could mean that the nuclei of barred galaxies are older than their circumnuclear regions. Finally, the stellar metallicities of barred and unbarred galaxies seem to be similar in their nuclear and circumnuclear regions.

\subsection{Circumnuclear spectra}

We performed an analysis of the circumnuclear region to understand the general features across a broader area, since the nuclear region analyzed here might be too small for us to make inferences regarding certain relations to the host galaxy. The goal was to carry out the same analysis as that applied to the nuclear region to evaluate if there is a connection between the nuclear and circumnuclear regions; otherwise,  we would assume they are distinguished regions with no continuity between them. 

\begin{figure*}
\begin{center}

  \includegraphics[scale=0.30]{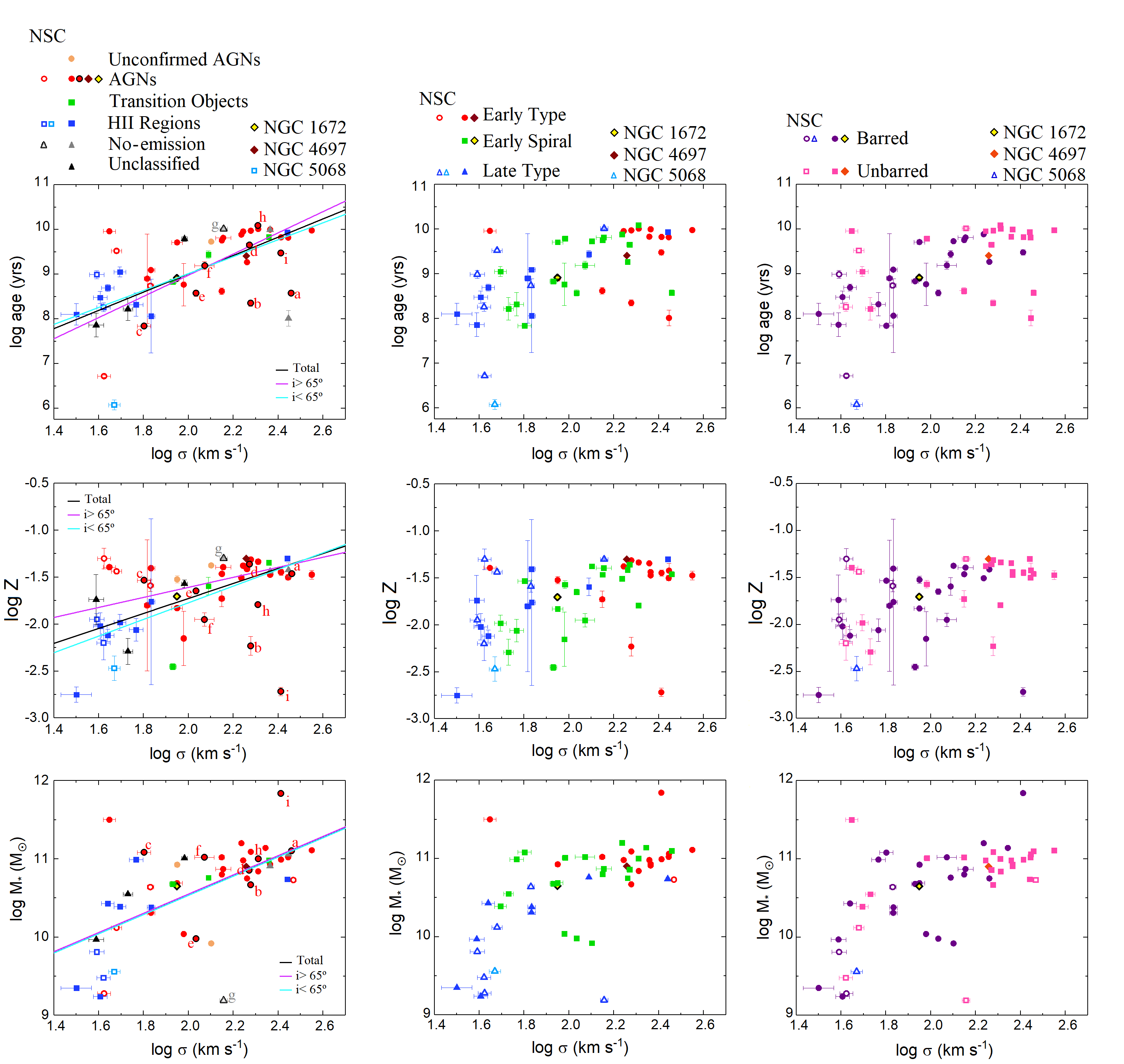}
  \caption{Age, metallicity, and total stellar mass versus the circumnuclear stellar velocity dispersion. Details are the same as in Fig.~\ref{sigma_plots_nuclear}.  \label{sigma_plots_circumnuclear}}
  
\end{center}
\end{figure*}

Figure~\ref{sigma_plots_circumnuclear} shows the plots of log(age) versus log($\sigma$), log(Z) versus log($\sigma$), and log($M_*$) versus log($\sigma$) for the extracted circumnuclear spectra. Correlations analogous to those observed for the nuclear spectra are detected here. In addition, the patterns regarding the presence or absence of an AGN, the morphological types, and the presence (or otherwise) of bars are also similar 
 to those seen in the nuclear spectra. Regarding the values we obtained (see Table~\ref{tableproperties}), the difference of $\sigma$ in the nuclear and circumnuclear regions ($|\sigma_{circumnuclear} - \sigma_{nuclear}|$) in barred galaxies is lower than the difference in unbarred galaxies. Additionally, 46\% of the barred galaxies have higher $\sigma$ values in the nucleus than in their circumnuclear region, against 56\% of the unbarred galaxies. From the 46\% with higher $\sigma$ values (barred galaxies), 62\% are confirmed AGNs; as for the unbarred galaxies, from the 56\% with high values in the nuclear region, we found that 79\% are confirmed AGNs.

The analysis of the correlation in the log(age) versus log($\sigma$) plots in the top panels of Fig.~\ref{sigma_plots_circumnuclear} reveals, for the starburst galaxies, that the stellar populations seem to be older in the circumnuclear region than in the nuclear region. On the other hand, the active galaxies seem to have a broader distribution of ages in the nuclear region than in the circumnuclear region. This means that in such cases, the circumnuclear region appears to be older than the nuclear region, when there is an AGN. By comparing the nuclear and circumnuclear results (Table ~\ref{tableproperties}), 61\% of the barred galaxies have older stellar populations in the nuclear region, rather than in the circumnuclear regions, compared to 52\% of the unbarred galaxies. Within these values, we see that, in the case of barred galaxies, 71\% contain AGNs and, for the unbarred galaxies, 69\% contain AGNs.

The correlations in the log(Z) versus log($\sigma$) plots in the middle panels of Fig.~\ref{sigma_plots_circumnuclear} show a larger scatter, with higher uncertainties, than the corresponding plots for the nuclear spectra. For starburst galaxies, the metallicity values are higher in the nuclear region. For AGNs and early spirals, the metallicities exhibit a broader range of values in the circumnuclear region than in the nuclear region. The circumnuclear regions of barred galaxies show lower metallicities than the nuclear regions in 79\% of galaxies. Within this sample, 59\% are confirmed AGNs. For the unbarred galaxies this proportion is 80\%, with 70\% of confirmed AGNs. Finally, the log($M_*$)  versus log($\sigma$) plots in the bottom panels of Fig.~\ref{sigma_plots_circumnuclear} show a remarkable similarity with the corresponding plots of the nuclear spectra.

When it comes to the nuclear activity, we observed that $\sim$ 63\% of confirmed AGNs have higher values of $\sigma$ and age in the nuclear region when compared with their circumnuclear region, while $\sim$ 84\% have higher values of metallicity in the nuclear region. However, most of the starburst galaxies have higher values of $\sigma$ and age in the circumnuclear region (80\%) and  they have also predominantly higher values of metallicity in the nuclear region (80\%) when compared with their circumnuclear regions.

Similarly to the behavior observed for the nuclear spectra, the linear fits reproduced the main patterns in the plots of Fig.~\ref{sigma_plots_circumnuclear}. The angular and the linear coefficients obtained for all galaxies, for galaxies with inclinations higher than 65\degr and for galaxies with inclinations lower than 65\degr, are compatible, at the 1-$\sigma$, 2-$\sigma$, or 3-$\sigma$ levels. These coefficients are also compatible (at the 1-$\sigma$, 2-$\sigma$, or 3-$\sigma$ levels) with those of the corresponding graphs of the nuclear spectra.

\begin{table*}[]
\centering
\caption{Coefficients of the linear fits in Fig.\ref{sigma_plots_circumnuclear}.}
{
\begin{tabular}{ccccccc}
\cline{2-7}
                       & \multicolumn{2}{c}{log (age) $\times$ log $\sigma$}                                                                          & \multicolumn{2}{c}{Z $\times$ log $\sigma$}                                                                                  & \multicolumn{2}{c}{log (M$_*$) $\times$ log $\sigma$}                                                                        \\ \cline{2-7} 
Sample                 & \begin{tabular}[c]{@{}c@{}}Angular\\ Coefficient\end{tabular} & \begin{tabular}[c]{@{}c@{}}Linear\\ Coefficient\end{tabular} & \begin{tabular}[c]{@{}c@{}}Angular\\ Coefficient\end{tabular} & \begin{tabular}[c]{@{}c@{}}Linear\\ Coefficient\end{tabular} & \begin{tabular}[c]{@{}c@{}}Angular\\ Coefficient\end{tabular} & \begin{tabular}[c]{@{}c@{}}Linear\\ Coefficient\end{tabular} \\ \hline
Total                  & 4.9 $\pm$ 0.6                                               & 2.1 $\pm$ 0.3                                               & -3.3 $\pm$ 0.3                                               & 0.80 $\pm$ 0.14                                              & 8.1 $\pm$ 0.5                                               & 1.23 $\pm$ 0.22                                               \\
i > 65$^{\circ}$ & 4.2 $\pm$ 1.0                                             & 2.4 $\pm$ 0.5                                               & -2.7 $\pm$ 0.5                                               & 0.53 $\pm$ 0.23                                                & 8.1 $\pm$ 0.9                                               & 1.2 $\pm$ 0.4                                               \\
i < 65$^{\circ}$    & 5.2 $\pm$ 0.8                                               & 1.9 $\pm$ 0.4                                                & -3.5 $\pm$ 0.4                                               & 0.89 $\pm$ 0.17                                              & 8.1 $\pm$ 0.5                                                 & 1.23 $\pm$ 0.26                                             \\ \hline
\end{tabular}
}
\label{coef_circumnuclear}
\end{table*}

\section{Discussion}\label{sec4}

The results in the previous sections indicate that there are clear correlations between $\sigma$ and age, metallicity, and $M_*$, both in the nuclear (dozens of pc) and circumnuclear regions (hundreds of pc) of the galaxies in our sample (although $M_*$ was estimated for the entire galaxies). Considering the correlation between $\sigma$ and $M_*$, we can also present these results as correlations between $M_*$ and the age and metallicity of the stellar populations in the nuclear and circumnuclear regions. The correlation involving the stellar metallicity is qualitatively consistent with the stellar mass-metallicity relation observed in many previous studies (e.g., \citealt{gal05,kir13,zah17,lia18,zha18}) and described in detail by \citet{mai19}. This correlation is also qualitatively consistent with the gas-phase mass-metallicity relation described in the literature (e.g., \citealt{hun12,lia18}). We tried to perform a quantitative comparison with the results from previous studies; however, we encountered many difficulties in our attempts to do so. Many of these studies did not provide linear fits, which would be required for a comparison. Others used definitions for the stellar metallicity different than ours (which is a light weighted mean). Considering all of these obstacles, we opted to not include such a comparison in this work.

The studies mentioned above were primarily based on long-slit spectra and, thus, they described correlations among the parameters obtained for large regions (on kiloparsec scales) of the galaxies. Some of the previous studies also considered only one type of galaxy; whereas in the present work, we included all types of galaxies in the correlation, also exploring their differences in morphological types, nuclear activity, presence of the bar, and inclinations. To the best of our knowledge, our analysis of these correlations is the first to be performed with data at such a high spatial resolution, so that the parameters involve specifically the nuclear regions of the galaxies. In other words, our results indicate that the stellar mass-metallicity relation is valid not only for large portions of the galaxies, but also specifically for the nuclear regions of galaxies. It is worth mentioning, however, that some studies involving CALIFA data (e.g., \citealt{zib22,zhu19}) and Multi-Unit Spectroscopic Explorer (MUSE) data \citep{pes21}; for example, having already established that the mass-metallicity relation holds locally inside galaxies. Similarly, the recent study of \citet{li25} detected similar correlations by analyzing the central spectra of the FoV of the IFU data from the MaNGA survey \citep{bun15}. However, the data used in these previous analyses have a spatial resolution significantly lower than that of our GMOS/IFU data. Nevertheless, these results support our findings.

Our results are not sufficient for us to evaluate any single most likely mechanism among the five presented in Section~\ref{sec1} (i.e., galactic outflows, different evolutionary stages, infall of metal-poor gas, IMF with a higher mass cutoff in massive galaxies, and a higher metallicity of the accreted gas in massive galaxies) to explain the correlation between $\sigma$ (and $M_*$) and stellar metallicity. To evaluate this expect, dedicated galactic evolution modeling (beyond the scope of this paper), including processes such as galactic outflows, gas infalls, and others would be required. It is worth mentioning that most of the AGNs in our sample, which often present intense outflows, are located in the high velocity dispersion-high metallicity end of the correlation we detected. Therefore, based on the first scenario to explain the mass-metallicity relation mentioned before, we conclude that outflows from these AGNs (in addition to possible galactic outflows) are not sufficient to remove metal-enriched gas from these galaxies, which are more massive and therefore have a deeper gravitational potential well. 

The fact that the parameters of the linear fits of the graphs obtained for the nuclear and circumnuclear regions were all compatible, at the 1-$\sigma$, 2-$\sigma,$ or 3-$\sigma$ levels (Tables \ref{coef_nuclear} and  \ref{coef_circumnuclear}), regardless of the inclination, might indicate that the correlations we have obtained for the central regions of the galaxies are not just similar, but actually the same as those obtained (in previous studies) for larger portions of the galaxies. We have to keep in mind, however, that our analyses were performed with a relatively small number of galaxies and, as a consequence, the uncertainties of certain parameters of the linear fits are somewhat high. However, this sample is representative of what we could expect for bigger samples in the Universe, since it comprises a large variety of objects and characteristics (i.e., the presence or absence of bars, morphologies, nuclear activities, evolution stages, etc.). 

Some of the obtained results, both for the nuclear and circumnuclear regions of the galaxies and regarding the division of the sample in barred and unbarred galaxies, must be discussed in further detail. To do so, we calculated the nuclear and circumnuclear average values of log(age), log(Z), and log($\sigma$) for barred and unbarred galaxies. The corresponding uncertainties of these average values were determined using a simple propagation of errors. The results are discussed below. These average values, however, must be taken with caution, as we did not take into account the statistical uncertainties, which are probably high due to our relatively small sample of objects. Another important remark is that such a comparison must take into account the morphological types of the galaxies. In our sample, the unbarred galaxies happen to have earlier morphological types, which implies in higher bulge masses and older stellar populations with higher metallicities. 

For the nuclear spectra, the average values of log($\sigma$) for barred and unbarred galaxies are log($\sigma_{nuc,~barred}$) = 1.909 $\pm$ 0.004 km s$^{-1}$ and log($\sigma_{nuc,~unbarred}$) = 2.174 $\pm$ 0.004 km s$^{-1}$, respectively. For the circumnuclear spectra, the average $\sigma$ values are log($\sigma_{circ,~barred}$) = 1.918 $\pm$ 0.005 km s$^{-1}$ and log($\sigma_{circ,~unbarred}$) = 2.197 $\pm$ 0.003 km s$^{-1}$. The results for barred and unbarred galaxies, both for the nuclear and circumnuclear regions, are not compatible, even at the 3-$\sigma$ level, with the barred galaxies values being lower than those of the unbarred galaxies. That, in principle, is in contradiction with predictions from simulations (e.g., \citealt{har14}), which state that barred galaxies should present higher central values of $\sigma$ than unbarred galaxies, due to the redistribution of angular momentum, leading to a mass increase and, therefore, to a stellar velocity dispersion increase in the central regions of the galaxies (where $\sigma$ is a tracer of the bulge mass). In our sample, the unbarred galaxies have earlier morphological types, which implies in higher bulge masses and, therefore, higher $\sigma$ values; even by removing the elliptical galaxies, we still obtain higher values of $\sigma$, compatible with the ones obtained with the whole unbarred sample. That means that the different morphological types that still are in the sample, even after the removal of elliptical galaxies, can still cause the higher values of $\sigma$. A possible approach to finding a solution would be a comparison with galaxies with the same morphological types and masses. However, our sample is too limited in number for a study of that scope.

The average values of log(age) for barred and unbarred galaxies are log(age)$_{nuc,~barred}$ = 8.96 $\pm$ 0.04 yrs and log(age)$_{nuc,~unbarred}$ = 9.410 $\pm$ 0.012 yrs, respectively, for the nuclear spectra, and log(age)$_{circ,~barred}$ = 8.86 $\pm$ 0.05 yrs and log(age)$_{circ,~unbarred}$ = 9.389 $\pm$ 0.015 yrs, for the circumnuclear spectra. Similarly to what was observed for the average $\sigma$ values, the results for barred and unbarred galaxies, both for the nuclear and circumnuclear regions, are not compatible, even at the 3-$\sigma$ level. In this case, we note that barred galaxies show younger nuclear and circumnuclear stellar populations than unbarred galaxies. That is possibly a consequence of feeding processes that happen under the influence of the bar and trigger star formation in those systems. However,  we must again take into account that the unbarred sample have a significant number of earlier type galaxies. Regardless of the presence or absence of an AGN, the bar has a strong influence on the stellar formation and enrichment of the nuclear region, since the bar usually induces transfer of gas and stars from the disk of the host galaxy to its nuclear region (e.g., \citealt{ku13,torquebars24}).

Regarding the average log(Z) values, for the nuclear spectra, we obtained log(Z)$_{nuc,~barred}$ = -1.64 $\pm$ 0.05 and log(Z)$_{nuc,~unbarred}$ = -1.566 $\pm$ 0.012, for the barred and unbarred galaxies, respectively, and, for the circumnuclear spectra, the calculated values were log(Z)$_{circ,~barred}$ = -1.78 $\pm$ 0.04 and log(Z)$_{circ,~unbarred}$ = -1.583 $\pm$ 0.012. For the nuclear spectra, the average values for barred and unbarred galaxies are compatible, at the 2-$\sigma$ level; whereas for the circumnuclear spectra, the values for barred and unbarred galaxies are not compatible, even at the 3-$\sigma$ level, with the unbarred galaxies being more metal-rich. These results are not consistent with those obtained by \citet{fra20}, using MaNGA data, who concluded that barred galaxies are older and more metal-rich than unbarred galaxies. This discrepancy may be related to the fact that the analysis of \citet{fra20} involved data with a spatial resolution much lower than our spatial resolution. In addition, we recall that we have to take into account the different morphologies in the sample of unbarred galaxies. Finally, besides analyzing the effects of the bar on the properties of the stellar populations in the central regions of the galaxies, previous studies compared the stellar populations along the bar and off-axis. For example, \citet{neu20} used MUSE data to reach the conclusion that intermediate-age stars ($\sim$ 2 - 6 Gyrs) shape a thinner part of the bar, while older stars (> 8 Gyrs) shape a rounder and thicker part of the bar. The authors also concluded that bars tend to be more metal-rich than the surrounding discs. \citet{neu24}, using MaNGA data, found that bars have higher stellar mass density and are more metal-rich than the discs at the same distance from the nucleus.

Regarding the correlation between $\sigma$ and ages of the stellar populations, \citet{smi09} found an analogous correlation for quiescent galaxies in the Shapley supercluster, using data with lower spatial resolution nonetheless. Such a correlation is certainly related to the fact that galaxies with the highest $\sigma$ (and, therefore, the highest $M_*$) are mostly early type galaxies, which are known to have smaller amounts of available gas, little star formation, and, as a consequence, older stellar populations. The relation between $\sigma$ and ages of the stellar populations has also been detected and discussed in detail for the Milky Way \citep{spi53,wie77,car85,hol09}. For external galaxies, \citet{wis15} observed an increase in the gas-phase $\sigma$ with redshift in disc galaxies, which suggests that the old stellar populations were formed from a kinematically hot gas and kept the high $\sigma$ values along time. Similarly to the case of the stellar mass-metallicity relation,  the main difference between our results and those of the previously cited studies is that we are analyzing the central regions of galaxies, while prior analyses were mostly based on lower spatial resolution data and therefore involved kiloparsec scales.

One interesting point to be discussed in the plots of log(age) versus log($\sigma$) in Fig.~\ref{sigma_plots_nuclear} is the presence of five AGNs (NGC 1566, NGC 1068, NGC 613, NGC 5128, and NGC 4594) somewhat below the observed correlation. This result can be described in the following way: for the range covered by the stellar velocity dispersions of these objects (1.93 < log($\sigma$) (km s$^{-1}$) < 2.40), their nuclear stellar populations is expected to be older than what is assumed based on observations. One explanation for this involves possible outflows from these AGNs. It is well known that AGN outflows can shut off star formation via a process called ``negative'' feedback (see \citealt{fab12} for a detailed review). However, there is also  so-called ``positive'' feedback, which involves AGN outflows triggering star formation \citep{mai17,gal19,mar22}. On this basis, we propose that the younger (than expected) nuclear stellar populations in these five AGNs could be an example of ``positive'' AGN feedback in nearby galaxies. For NGC 1566 and NGC 613, the bar may play an important role in triggering star formation in their rich circumnuclear environment (e.g., \citealt{,daSilva2017,613a,613b}). There is also the possibility that the AGN outflows increased the $\sigma$ values of the star-forming gas. In that case, the recently formed stars kept these high $\sigma$ values. It is worth pointing out that two additional objects seem considerably far from the correlation: NGC 300 and NGC 1365, indicated in the middle panel of Fig.~\ref{sigma_plots_nuclear}. In the case of NGC 300, \citet{Fielder25} support the evidence of an accretion event; in the case of NGC 1365, the predominancy of old stellar populations can be caused by AGN and supernova feedback in its nuclear region \citep{Sextl24}. NGC 7213 is an outlier in the plot of log(Z) versus log($\sigma$) in Fig.~\ref{sigma_plots_nuclear}. For its $\sigma$ value, the metallicity should be higher than what has been observed. Since NGC 7213 is an AGN, a possible explanation is that AGN outflows may have expelled high-metallicity gas from the nuclear region, resulting in the observed low metallicity values. NGC 1316 can be seen as an outlier in the plot of log($M_*$)  versus log($\sigma$); however, the reason for that behavior is still unclear.

Nine galaxies in our sample (GC 247, NGC 300,
NGC 3115, NGC 3621, NGC 5068, NGC 5102, NGC 5236,
NGC 7090, and NGC 7793) have nuclear star clusters, according to \citet{hoyer2021}. They can have very different stellar population properties, compared to the underlying host galaxy \citep{fah21}. However, we note that the majority of these nine galaxies follow the main relations we observe in Figs.~\ref{sigma_plots_nuclear} and \ref{sigma_plots_circumnuclear}. Therefore, we conclude that the presence of nuclear star clusters in the galaxies is not related to the existence of outliers in the relations we have observed in this study.

In general, we have to take into account that most of confirmed AGNs present higher values of metallicity, stellar velocity dispersion, and ages in the nucleus when compared with their circumnuclear region. In terms of metallicity, starburst galaxies also have higher metallicities in their nuclear region when compared with their circumnuclear region. That means that the presence of AGNs is not essential to the enrichment of the circumnuclear region, which is almost equal in both samples (only in 26\% of confirmed AGNs and in 20\% of starbursts, the circumnuclear region has a higher metallicity than the nucleus). We also have to take into consideration the fact that the circumnuclear regions are older and have higher stellar velocity dispersion in 80\% of starburst galaxies. This demonstrates the tendency for AGNs to be surrounded by a younger stellar environment, which could indicate the positive feedback influence of AGNs in the circumnuclear regions; alternatively, AGNs are more predominant in galaxies where the processes of feeding are efficient enough to also spur star formation in the circumnuclear region. We also found out (as shown in Fig.~\ref{metal_age_nuc}) that we can separate active galaxies and starburst by looking at the mean stellar populations age of their nuclear spectra. This plot suggests that starburst galaxies do not have average stellar populations more than one billion years old. On the other hand, most of the outliers (and galaxies with NSC) have younger stellar populations than most of the AGNs. Predominantly, AGNs are accompanied (in the nuclear region) by older stellar populations with high metallicites (independent of the bar presence), as stated above. By observing this plot, we can see that all ``unconfirmed AGNs" and at least one transition object (NGC 3585) of the sample are in the area of AGNs, suggesting that those objects might be AGNs as well.

Future studies with larger samples using high spatial resolution data are necessary to confirm the trends we have observed and detailed in this study. However, since this sample has no bias, we believe that these correlations and trends, which are consistent with the correlations we observe in the scenarios for each individual object, will prevail.

\section{Conclusions}\label{sec5}

We analyzed the relations between stellar velocity dispersion ($\sigma$), stellar populations age, stellar metallicity, and total stellar mass of the host galaxy ($M_*$) in the nuclear (dozens of parsecs) and circumnuclear (hundreds of parsecs) regions of nearby galaxies. We based our study on a sub-sample of the DIVING$^{3D}$ survey, cited a Paper I. The main conclusions of this work are listed below. 
\begin{itemize}

\item We determined valid correlations between $\sigma$ and  age, metallicity, and total stellar mass for both the nuclear and circumnuclear regions of the galaxies in our sample.

\item The correlation between $\sigma$ and the stellar metallicity is qualitatively consistent with the stellar mass-metallicity relation described in previous studies. However, we have shown,  for the first time, that this correlation is also valid for the central regions of galaxies (and not only for large portions of the galaxies) independently of the morphological types.

\item The parameters of the linear fits of the graphs obtained for the nuclear and circumnuclear regions were all compatible, at the 1-$\sigma$, 2-$\sigma$, or 3-$\sigma$ levels, independent of inclinations, including different morphological types, nuclear activity, presence of bars, and NSCs. This suggests that the correlations are the same for the scales we studied and for larger scales within the galaxies; therefore, they can be generalized for all galaxies. 

\item Based on our sample, we observe that unbarred galaxies tend to have older stellar populations, higher stellar metallicities, and higher stellar velocity dispersion than barred galaxies, in their central regions. This may be a consequence of the influence of the bar, since it usually transfers gas and stars from the disk to the central regions. However, we must also take into account that the sample of unbarred galaxies contains earlier morphological types, which results in the higher values observed. Further studies, with larger samples are required to establish comparisons between barred and unbarred galaxies, with the same mass and morphological type.

\item For starburst galaxies, the mean ages of their stellar populations  are not higher than one billion years old. Active galaxies  predominantly have high-metallicity, older stellar populations in the nucleus, while AGNs with NSCs show a more diverse range of mean ages.

\item Some AGNs are positioned above the main correlation in the plot of nuclear log(age) versus log($\sigma$), suggesting that their nuclear stellar populations are younger than expected, considering their nuclear $\sigma$ values. This may be a consequence of  ``positive'' AGN feedback, triggering star formation in the nuclear regions of these objects. Another fact that supports the notion of positive feedback is the predominancy of AGNs in younger circumnuclear regions versus starburst galaxies, which, in turn, predominantly exhibit a younger nucleus when compared to their circumnuclear regions.

\end{itemize}

\begin{acknowledgements}

PdS thanks Funda\c{c}\~ao de Amparo \`a Pesquisa do Estado de S\~ao Paulo (FAPESP) for the support under grants 2020/13315-3 and 2022/14382-1;  
RBM and TVR thank Conselho Nacional Cient\'ifico e Tecnol\'ogico (CNPq) for the support under grants 309976/2022-7 and 304584/2022-3, respectively.    
FP acknowledges support from the Horizon Europe research and innovation programme under the Maria Skłodowska-Curie grant “TraNSLate” No 101108180, and from the Agencia Estatal de Investigación del Ministerio de Ciencia e Innovación (MCIN/AEI/10.13039/501100011033) under grant (PID2021-128131NB-I00) and the European Regional Development Fund (ERDF) ``A way of making Europe''.
BB acknowledges partial financial support from FAPESP, CNPq and Coordena\c{c}\~ao de Aperfei\c{c}oamento de Pessoal de N\'ivel Superior (CAPES).
This work was based on observations obtained at the Gemini Observatory (processed using the Gemini IRAF package), which is operated by the
Association of Universities for Research in Astronomy, Inc., under
a cooperative agreement with the NSF on behalf of the Gemini partnership: the National Science Foundation (United States),
the National Research Council (Canada), CONICYT (Chile), the
Australian Research Council (Australia), Minist\'erio da Ci\^encia, Tecnologia e Inova\c{c}\~ao (Brazil) and Ministerio de Ciencia, Tecnolog\'ia e Innovaci\'on Productiva (Argentina). This research has made
use of the NASA/IPAC Extragalactic Database (NED), which is
operated by the Jet Propulsion Laboratory, California Institute of
Technology, under contract with the National Aeronautics and Space
Administration. We also acknowledge the usage of the HyperLeda
data base (http://leda.univ-lyon1.fr). 

\end{acknowledgements}

%
\bibliographystyle{aa} 
\bibliography{references} 
%

\begin{appendix}

\section{Galaxy sample properties}\label{app_table}

Table \ref{table1} lists the sample and their main properties taken from previous studies, and the size of the FoV of the data cubes of each object analyzed in this work.

\begin{table*}[h!]
\centering
\caption{Properties of the sample of galaxies.}\label{table1}
\resizebox{0.9\textwidth}{!}{
\begin{tabular}{cccccccc}
\hline
Galaxy   & Morphological Type & Distance (Mpc) & Inclination (deg) & log M* (Msun) & Nuclear Activity & r (pc) & FOV size (pc) \\ \hline
NGC 134  & SAB(s)bc           & 18.3$^a$       & 90                      & 9.92          & L                  & 20.41  & 266 $\times$ 395 \\
NGC 157  & SAB(rs)bc          & 12.9$^b$       & 61.8                       & 10.99         & T                 & 10.63  & 175 $\times$ 275 \\
NGC 247  & SAB(s)d            & 3.34$^c$       &  76.4                  & 9.28          & HII/T/S          & 3.89   & 47 $\times$ 72   \\
NGC 253  & SAB(s)c            & 3.05$^c$       & 90                    & 10.76         & T                  & 3.40   & 41 $\times$ 63   \\
NGC 300  & SA(s)d             & 2.17$^a$       & 48.5                      & 9.19          & --            & 2.52   & 34 $\times$ 49   \\
NGC 613  & SB(rs)bc           & 21.4$^d$          & 35.7                    & 11.08         & L                 & 25.94  & 311 $\times$ 394 \\
NGC 720  & E5                 & 30.7$^e$         & 90                  & 11.02         & HII/T/L/S        & 95.26  & 432 $\times$ 707 \\
NGC 908  & SA(s)c             & 18.5$^a$             & 64.7                    & 10.74         & HII              & 25.11  & 287 $\times$ 377 \\
NGC 936  & SB0                & 19.9$^f$            & 50.4              & 10.926$^{1}$  & L                & 29.91  & 251 $\times$ 410 \\
NGC 1068 & (R)SA(rs)b         & 10.1$^d$             & 34.7                  & 10.861$^{1}$  & S                  & 17.63  & 135 $\times$ 235 \\
NGC 1097 & SB(s)b             & 15.4$^a$              & 54.8                   & 10.87         & L/S               & 17.17  & 202 $\times$ 317 \\
NGC 1187 & SB(r)c             & 21.4$^a$            & 44.1                 & 10.38         & HII            & 29.05  & 311 $\times$ 451 \\
NGC 1232 & SAB(rs)c           & 14.5$^d$            &   32.7            & 10.38         & --       & 58.35  & 928 $\times$ 422 \\
NGC 1291 & (R)SB0/a(s)        & 8.6$^g$          & 29.3                        & 10.75         & L                 & 10.42  & 100 $\times$ 167 \\
NGC 1300 & SB(rs)bc           & 14.5$^d$               & 61.8                  & 10.3$^{2}$    & L               & 14.06  & 218 $\times$ 323 \\
NGC 1313 & SB(s)d             & 3.7$^g$               & 34.8               & 9.35          & HII             & 4.84   & 55 $\times$ 75   \\
NGC 1316 & SAB0               & 20.8$^h$              & 67.8             & 11.84$^{3}$   & L/S               & 37.31  & 313 $\times$ 439 \\
NGC 1365 & SB(s)b             & 13.6$^a$              & 62.6                  & 11.02         & S                  & 18.46  & 181 $\times$ 254 \\
NGC 1380 & SA0                & 20.6$^i$             & 90            & 10.93         & L                & 49.94  & 295 $\times$ 444 \\
NGC 1395 & E2                 & 24.3$^f$                 & 47.4                & 11.02         & HII/T/L/S        & 49.48  & 359 $\times$ 501 \\
NGC 1398 & (R’)SB(r)ab        & 28.6$^a$               & 47.5          & 11.14         & L               & 24.96  & 416 $\times$ 610 \\
NGC 1399 & E1                 & 21.1$^i$                 & 34.2             & 11.11         & HII/T/S            & 48.08  & 286 $\times$ 455 \\
NGC 1404 & E1                 & 19$^e$                    & 44.7           & 10.99$^{4}$   & HII/T/S          & 32.24  & 258 $\times$ 428 \\
NGC 1407 & E0                 & 25.1$^j$                    & 30.3          & 11.21         & --                & 43.81  & 341 $\times$ 578 \\
NGC 1433 & (R’)SB(r)ab        & 8.32$^c$                     & 67.4      & 10.04         & L/S             & 10.49  & 119 $\times$ 169 \\
NGC 1549 & E0-1               & 16.3$^e$                  & 42.3      & 10.91         & --               & 25.29  & 237 $\times$ 332 \\
NGC 1553 & SA0                & 9.51$^e$                 & 56.6         & 11.09         & L/S                & 13.83  & 104 $\times$ 166 \\
NGC 1559 & SB(s)cd            & 11.6$^a$                & 59.9          & 10.43         & HII             & 11.81  & 163 $\times$ 236 \\
NGC 1566 & SAB(s)bc           & 18$^k$                & 49.1              & 9.98          & S                 & 30.54  & 275 $\times$ 358 \\
NGC 1574 & SA0                & 20.1$^f$               & 76.6              & 10.84         & L/S           & 38.98  & 283 $\times$ 419 \\
NGC 1672 & SB(s)b             & 11.4$^l$                 & 28.9                 & 10.65         & S                & 11.05  & 133 $\times$ 232 \\
NGC 1792 & SA(rs)bc           & 10.8$^a$                & 63.1          & 10.39         & HII              & 13.09  & 152 $\times$ 220 \\
NGC 1808 & (R)SAB(s)a         & 9.51$^a$                 & 82.7          & 10.68         & T                 & 10.14  & 122 $\times$ 191 \\
NGC 2442 & SAB(s)bc           & 20.1$^m$                 & 50.3          & 10.8          & L                 & 29.23  & 287 $\times$ 390 \\
NGC 2835 & SB(rs)c            & 8.75$^n$                 & 56.2          & 9.97          & HII/T/L/S        & 7.64   & 121 $\times$ 163 \\
NGC 2997 & SAB(rs)c           & 13.1$^d$               & 53.7          & 10.31         & L                & 16.51  & 187 $\times$ 273 \\
NGC 3115 & S0                 & 9.68$^o$              & 90                  & 10.73         & T/L/S             & 15.49  & 106 $\times$ 209 \\
NGC 3585 & E6                 & 13.1$^e$              & 90                  & 10.98         & T               & 18.42  & 197 $\times$ 289 \\
NGC 3621 & SA(s)d             & 6.4$^a$              & 67.6                   & 10.12         & L/S            & 6.21   & 92 $\times$ 140  \\
NGC 3923 & E4-5               & 16$^e$               & 90                  & 11.06         & --               & 27.93  & 240 $\times$ 345 \\
NGC 4030 & SA(s)bc            & 29.9$^n$              & 47                  & 11.01         & HII/L             & 37.69  & 435 $\times$ 623 \\
NGC 4594 & SA(s)a             & 21.7$^p$             & 59.4                & 11.1          & L                & 37.87  & 316 $\times$ 500 \\
NGC 4697 & E6                 & 12.4$^n$              & 90               & 10.9$^{3}$    & L/S              & 13.23  & 150 $\times$ 246 \\
NGC 4699 & SAB(rs)b           & 19.5$^a$             & 42.1                & 11.2          & L/S               & 30.25  & 279 $\times$ 425 \\
NGC 4753 & I0                 & 24.1$^n$              & 78.2                   & 10.98         & L/S              & 26.87  & 310 $\times$ 508 \\
NGC 5068 & SAB(rs)cd          & 6.7$^g$               & 27.3              & 9.56          & HII             & 4.22   & 94 $\times$ 133  \\
NGC 5102 & SA0                & 4$^o$                 & 90             & 9.48          & HII/T/L/S          & 4.07   & 59 $\times$ 86   \\
NGC 5128 & S0                 & 3.53$^q$              & 45.3              & 10.67         & L/S              & 22.25  & 488 $\times$ 770 \\
NGC 5236 & SAB(s)c            & 4.79$^r$              & 15.3             & 10.64         & HII/L           & 5.11   & 63 $\times$ 101  \\
NGC 5247 & SA(s)bc            & 22.2$^g$              & 38.5            & 10.55         & HII/T/L/S       & 20.45  & 344 $\times$ 457 \\
NGC 5643 & SAB(rs)c           & 16.9$^g$              & 29.6                 & --            & S                  & 16.39  & 254 $\times$ 348 \\
NGC 6744 & SAB(r)bc           & 7.66$^a$              & 53.5            & 10.69         & L               & 6.31   & 117 $\times$ 150 \\
NGC 7090 & SBc                & 6.22$^a$              & 90            & 9.81          & HII               & 6.63   & 90 $\times$ 122  \\
NGC 7213 & SA(s)a             & 22$^g$               & 39.1                & 11            & L                & 37.33  & 251 $\times$ 363 \\
NGC 7424 & SAB(rs)cd          & 11.5$^g$             & 59              & 9.24          & HII               & 10.59  & 176 $\times$ 237 \\
NGC 7793 & SA(s)d             & 3.73$^a$             & 63.5                & 9.48          & HII               & 3.80   & 58 $\times$ 80   \\
IC 1459  & E3-4             & 28.7$^n$              & 74              & 11.5$^{3}$    & L                  & 54.27  & 404 $\times$ 633 \\  \hline
\end{tabular}}
\\
\scriptsize{\textit{Notes.} The galaxies morphological types were taken from \citet{RC3}. The distances of the galaxies were taken from: (a) \citet{Tully16}, (b) \citet{Erwin17}, (c) \citet{Tully09}, (d) \citet{Nasonova11}, (e) \citet{Springob14}, (f) \citet{Blakeslee01}, (g) \citet{Tully88}, (h) \citet{Cantiello13}, (i) \citet{Blakeslee10}, (j) \citet{Cantiello05}, (k) \citet{Sabbi18}, (l) \citet{Bottinelli86}, (m) \citet{Riess16}, (n) \citet{Tully13}, (o) \citet{Tonry01}, (p) \cite{Sorce14}, (q) \citet{Theureau07}, and (r) \citet{Pierce94}. The total stellar mass were mainly taken from \citet{biteau21}, and the others were taken from (1) \citet{Sheth10}, (2) \citet{Davis19}, (3) \citet{Greene20}, and (4) \citet{harris20}. The nuclear activity was taken from Paper I \citep{men22}, L states for Low Ionization Nuclear Emission-Line Regions (LINERs), HII states for HII regions, T for transition objects, and S for Seyferts. The radius (r), for which the nuclear spectra were obtained, is also shown in pc, together with the x and y axis of the FoVs of each data cube. The galaxies inclinations were taken from Hyperleda \citep{hyperleda}.}

\end{table*}

Table \ref{tableproperties} shows the parameters, resulting from the spectral synthesis, that we used in all plots of this work, and also to estimate the percentages in the results and discussion sections. The nuclear emission here refers to the final nuclear emission classifications we used to separate the sample, based on their nuclear activity studied in previous works. 

\begin{table*}[h!]
\centering
\caption{Properties and results obtained for the galaxies used in this work.}\label{tableproperties}
\resizebox{0.9\textwidth}{!}{%
\begin{tabular}{cccccccc}
\hline
\multirow{2}{*}{Galaxy} & \multirow{2}{*}{\begin{tabular}[c]{@{}c@{}}Nuclear \\ Emission\end{tabular}} & \multicolumn{2}{c}{Log age (yrs)}        & \multicolumn{2}{c}{Log Z}               & \multicolumn{2}{c}{\begin{tabular}[c]{@{}c@{}}Log $\sigma$ \\ (km/s)\end{tabular}} \\ \cline{3-8} 
                        &                                                                              & Nuclear             & Circumnuclear      & Nuclear            & Circumnuclear      & Nuclear                                  & Circumnuclear                           \\ \hline
NGC 134                 & Unconfirmed AGN                                                            & 9.818 $\pm$ 0.025   & 9.729 $\pm$ 0.023  & -1.31 $\pm$ 0.06   & -1.375 $\pm$ 0.018 & 2.015 $\pm$ 0.010                        & 2.101 $\pm$ 0.003                       \\
NGC 936                 & Unconfirmed AGN                                                            & 9.771 $\pm$ 0.015   & 8.90 $\pm$ 0.06    & -1.42 $\pm$ 0.03   & -1.52 $\pm$ 0.04   & 2.3276 $\pm$ 0.0020                      & 1.950 $\pm$ 0.010                       \\
NGC 1232                & Unconfirmed AGN                                                            & 10.229 $\pm$ 0.022  & ----               & -1.94 $\pm$ 0.05   & ----               & 2.001 $\pm$ 0.007                        & ----                                    \\
NGC 247                 & AGN                                                                          & 7.91 $\pm$ 0.15     & 6.720 $\pm$ 0.024  & -2.39 $\pm$ 0.17   & -1.30 $\pm$ 0.11   & 1.636 $\pm$ 0.028                        & 1.624 $\pm$ 0.029                       \\
NGC 613                 & AGN                                                                          & 7.90 $\pm$ 0.03     & 9.48 $\pm$ 0.06    & -1.54 $\pm$ 0.04   & -2.72 $\pm$ 0.05   & 2.00 $\pm$ 0.04                          & 1.80 $\pm$ 0.03                         \\
NGC 720                 & AGN                                                                          & 10.013 $\pm$ 0.005  & 8.62 $\pm$ 0.07    & -1.30 $\pm$ 0.05   & -1.73 $\pm$ 0.09   & 2.4679 $\pm$ 0.0013                      & 2.148 $\pm$ 0.012                       \\
NGC 1068                & AGN                                                                          & 7.35 $\pm$ 0.07     & 9.998 $\pm$ 0.020  & -1.422 $\pm$ 0.015 & -1.473 $\pm$ 0.010 & 2.086 $\pm$ 0.016                        & 2.270 $\pm$ 0.013                       \\
NGC 1097                & AGN                                                                          & 9.69 $\pm$ 0.13     & 8.8 $\pm$ 0.5      & -1.31 $\pm$ 0.05   & -2.15 $\pm$ 0.29   & 2.264 $\pm$ 0.005                        & 2.15 $\pm$ 0.03                         \\
NGC 1291                & AGN                                                                          & 10.088 $\pm$ 0.019  & 10.01 $\pm$ 0.03   & -1.301 $\pm$ 0.025 & -1.335 $\pm$ 0.026 & 2.2292 $\pm$ 0.0028                      & 2.2614 $\pm$ 0.0023                     \\
NGC 1300                & AGN                                                                          & 9.377 $\pm$ 0.029   & ----               & -1.443 $\pm$ 0.029 & ----               & 1.956 $\pm$ 0.009                        & ----                                    \\
NGC 1316                & AGN                                                                          & 9.699 $\pm$ 0.010   & 9.52 $\pm$ 0.03    & -1.382 $\pm$ 0.013 & -1.438 $\pm$ 0.025 & 2.3602 $\pm$ 0.0016                      & 2.4118 $\pm$ 0.0014                     \\
NGC 1365                & AGN                                                                          & 10.26 $\pm$ 0.05    & 9.408 $\pm$ 0.027  & -1.30 $\pm$ 0.05   & -1.301 $\pm$ 0.005 & 1.57 $\pm$ 0.05                          & 2.07 $\pm$ 0.04                         \\
NGC 1380                & AGN                                                                          & 9.913 $\pm$ 0.018   & 9.956 $\pm$ 0.021  & -1.301 $\pm$ 0.005 & -1.378 $\pm$ 0.016 & 2.3741 $\pm$ 0.0018                      & 2.3645 $\pm$ 0.0019                     \\
NGC 1395                & AGN                                                                          & 9.857 $\pm$ 0.011   & 9.944 $\pm$ 0.010  & -1.333 $\pm$ 0.011 & -1.368 $\pm$ 0.016 & 2.4616 $\pm$ 0.0027                      & 2.4448 $\pm$ 0.0010                     \\
NGC 1398                & AGN                                                                          & 10.204 $\pm$ 0.006  & 9.962 $\pm$ 0.012  & -1.328 $\pm$ 0.008 & -1.393 $\pm$ 0.015 & 2.3021 $\pm$ 0.0019                      & 2.3436 $\pm$ 0.0017                     \\
NGC 1399                & AGN                                                                          & 9.872 $\pm$ 0.012   & 9.98 $\pm$ 0.03    & -1.377 $\pm$ 0.019 & -1.47 $\pm$ 0.05   & 2.6012 $\pm$ 0.0010                      & 2.5501 $\pm$ 0.0010                     \\
NGC 1404                & AGN                                                                          & 9.918 $\pm$ 0.010   & 9.825 $\pm$ 0.016  & -1.344 $\pm$ 0.013 & -1.45 $\pm$ 0.03   & 2.3952 $\pm$ 0.0016                      & 2.4131 $\pm$ 0.0014                     \\
NGC 1433                & AGN                                                                          & 9.33 $\pm$ 0.03     & 9.268 $\pm$ 0.019  & -1.413 $\pm$ 0.017 & -1.413 $\pm$ 0.025 & 1.996 $\pm$ 0.007                        & 1.979 $\pm$ 0.008                       \\
NGC 1553                & AGN                                                                          & 10.015 $\pm$ 0.018  & 9.973 $\pm$ 0.010  & -1.33 $\pm$ 0.04   & -1.313 $\pm$ 0.008 & 2.2938 $\pm$ 0.0026                      & 2.2782 $\pm$ 0.0021                     \\
NGC 1566                & AGN                                                                          & 6.87 $\pm$ 0.13     & 8.57 $\pm$ 0.07    & -1.39 $\pm$ 0.23   & -1.65 $\pm$ 0.03   & 1.929 $\pm$ 0.009                        & 2.033 $\pm$ 0.005                       \\
NGC 1574                & AGN                                                                          & 9.852 $\pm$ 0.013   & 9.975 $\pm$ 0.011  & -1.30 $\pm$ 0.05   & -1.338 $\pm$ 0.011 & 2.4209 $\pm$ 0.0015                      & 2.3110 $\pm$ 0.0018                     \\
NGC 1672                & AGN                                                                          & 9.00 $\pm$ 0.04     & 8.92 $\pm$ 0.03    & -1.56 $\pm$ 0.04   & -1.704 $\pm$ 0.027 & 2.002 $\pm$ 0.007                        & 1.949 $\pm$ 0.009                       \\
NGC 2442                & AGN                                                                          & 9.71 $\pm$ 0.07     & 9.76 $\pm$ 0.04    & -1.33 $\pm$ 0.05   & -1.465 $\pm$ 0.018 & 2.130 $\pm$ 0.006                        & 2.149 $\pm$ 0.003                       \\
NGC 2997                & AGN                                                                          & 9.64 $\pm$ 0.06     & 9.093 $\pm$ 0.023  & -1.44 $\pm$ 0.04   & -1.404 $\pm$ 0.022 & 1.863 $\pm$ 0.021                        & 1.833 $\pm$ 0.017                       \\
NGC 3115                & AGN                                                                          & 9.928 $\pm$ 0.009   & 9.655 $\pm$ 0.014  & -1.301 $\pm$ 0.008 & -1.358 $\pm$ 0.015 & 2.5134 $\pm$ 0.0020                      & 2.4681 $\pm$ 0.0016                     \\
NGC 3621                & AGN                                                                          & 9.20 $\pm$ 0.09     & 9.19 $\pm$ 0.09    & -1.87 $\pm$ 0.04   & -1.95 $\pm$ 0.07   & 1.684 $\pm$ 0.027                        & 1.679 $\pm$ 0.023                       \\
NGC 4594                & AGN                                                                          & 8.94 $\pm$ 0.04     & 8.58 $\pm$ 0.06    & -1.422 $\pm$ 0.009 & -1.46 $\pm$ 0.03   & 2.4023 $\pm$ 0.0014                      & 2.4586 $\pm$ 0.0011                     \\
NGC 4697                & AGN                                                                          & 10.147 $\pm$ 0.022  & 9.814 $\pm$ 0.023  & -1.337 $\pm$ 0.016 & -1.393 $\pm$ 0.020 & 2.291 $\pm$ 0.003                        & 2.2579 $\pm$ 0.0024                     \\
NGC 4699                & AGN                                                                          & 9.974 $\pm$ 0.017   & 9.887 $\pm$ 0.017  & -1.448 $\pm$ 0.015 & -1.507 $\pm$ 0.016 & 2.2643 $\pm$ 0.0028                      & 2.236 $\pm$ 0.003                       \\
NGC 4753                & AGN                                                                          & 9.997 $\pm$ 0.021   & 9.819 $\pm$ 0.026  & -1.38 $\pm$ 0.03   & -1.501 $\pm$ 0.027 & 2.180 $\pm$ 0.003                        & 2.244 $\pm$ 0.003                       \\
NGC 5102                & AGN                                                                          & 8.498 $\pm$ 0.010   & ----               & -2.186 $\pm$ 0.029 & ----               & 1.71 $\pm$ 0.03                          & ----                                    \\
NGC 5128                & AGN                                                                          & 8.22 $\pm$ 0.10     & 8.35 $\pm$ 0.07    & -1.98 $\pm$ 0.14   & -2.23 $\pm$ 0.10   & 2.111 $\pm$ 0.014                        & 2.278 $\pm$ 0.005                       \\
NGC 5236                & AGN                                                                          & 9.37 $\pm$ 0.04     & 8.74 $\pm$ 0.05    & -1.47 $\pm$ 0.05   & -1.59 $\pm$ 0.06   & 1.863 $\pm$ 0.011                        & 1.830 $\pm$ 0.015                       \\
NGC 5643                & AGN                                                                          & 8.6 $\pm$ 0.9       & 8.9 $\pm$ 1.0      & -1.5 $\pm$ 1.2     & -1.8 $\pm$ 0.7     & 1.848 $\pm$ 0.011                        & 1.816 $\pm$ 0.014                       \\
NGC 6744                & AGN                                                                          & 9.86 $\pm$ 0.04     & 9.71 $\pm$ 0.05    & -1.53 $\pm$ 0.05   & -1.829 $\pm$ 0.017 & 1.823 $\pm$ 0.017                        & 1.949 $\pm$ 0.024                       \\
NGC 7213                & AGN                                                                          & 9.89 $\pm$ 0.05     & 10.091 $\pm$ 0.014 & -2.23 $\pm$ 0.03   & -1.793 $\pm$ 0.027 & 2.32 $\pm$ 0.04                          & 2.3098 $\pm$ 0.0023                     \\
IC 1459                 & AGN                                                                          & 9.999 $\pm$ 0.019   & 7.84 $\pm$ 0.05    & -1.43 $\pm$ 0.04   & -1.533 $\pm$ 0.025 & 2.5513 $\pm$ 0.0007                      & 1.647 $\pm$ 0.027                       \\
NGC 253                 & Transition Object                                                            & 9.55 $\pm$ 0.14     & 9.44 $\pm$ 0.08    & -1.68 $\pm$ 0.05   & -1.59 $\pm$ 0.09   & 1.986 $\pm$ 0.028                        & 2.089 $\pm$ 0.005                       \\
NGC 1808                & Transition Object                                                            & 8.60 $\pm$ 0.08     & 8.84 $\pm$ 0.04    & -2.36 $\pm$ 0.09   & -2.45 $\pm$ 0.03   & 1.80 $\pm$ 0.04                          & 1.930 $\pm$ 0.015                       \\
NGC 3585                & Transition Object                                                            & 9.9356 $\pm$ 0.0027 & 9.830 $\pm$ 0.014  & -1.301 $\pm$ 0.05  & -1.344 $\pm$ 0.018 & 2.371 $\pm$ 0.007                        & 2.359 $\pm$ 0.003                       \\
NGC 157                 & HII Region                                                                   & 8.36 $\pm$ 0.12     & 8.99 $\pm$ 0.09    & -1.572 $\pm$ 0.048 & -1.95 $\pm$ 0.06   & 1.757 $\pm$ 0.021                        & 1.77 $\pm$ 0.03                         \\
NGC 908                 & HII Region                                                                   & 8.09 $\pm$ 0.09     & 9.935 $\pm$ 0.023  & -1.499 $\pm$ 0.051 & -1.301 $\pm$ 0.016 & 1.823 $\pm$ 0.019                        & 2.441 $\pm$ 0.007                       \\
NGC 1187                & HII Region                                                                   & 8.17 $\pm$ 0.08     & 8.1 $\pm$ 0.8      & -2.22 $\pm$ 0.09   & -1.8 $\pm$ 0.9     & 1.807 $\pm$ 0.022                        & 1.834 $\pm$ 0.017                       \\
NGC 1313                & HII Region                                                                   & 7.27 $\pm$ 0.11     & 8.10 $\pm$ 0.24    & -2.55 $\pm$ 0.16   & -2.75 $\pm$ 0.08   & 1.55 $\pm$ 0.05                          & 1.50 $\pm$ 0.07                         \\
NGC 1559                & HII Region                                                                   & 8.48 $\pm$ 0.25     & 8.70 $\pm$ 0.08    & -1.89 $\pm$ 0.17   & -2.12 $\pm$ 0.07   & 1.59 $\pm$ 0.03                          & 1.640 $\pm$ 0.028                       \\
NGC 1792                & HII Region                                                                   & 8.94 $\pm$ 0.08     & 9.05 $\pm$ 0.11    & -1.84 $\pm$ 0.07   & -1.98 $\pm$ 0.08   & 1.655 $\pm$ 0.026                        & 1.694 $\pm$ 0.026                       \\
NGC 5068                & HII Region                                                                   & 7.75 $\pm$ 0.06     & 6.08 $\pm$ 0.11    & -1.97 $\pm$ 0.21   & -2.47 $\pm$ 0.13   & 1.62 $\pm$ 0.03                          & 1.669 $\pm$ 0.025                       \\
NGC 7090                & HII Region                                                                   & 7.96 $\pm$ 0.28     & 8.32 $\pm$ 0.26    & -1.71 $\pm$ 0.23   & -2.06 $\pm$ 0.12   & 1.59 $\pm$ 0.03                          & 1.59 $\pm$ 0.03                         \\
NGC 7424                & HII Region                                                                   & 8.03 $\pm$ 0.25     & 8.48 $\pm$ 0.14    & -1.5 $\pm$ 0.3     & -2.02 $\pm$ 0.14   & 1.59 $\pm$ 0.03                          & 1.61 $\pm$ 0.03                         \\
NGC 7793                & HII Region                                                                   & 8.26 $\pm$ 0.05     & 8.26 $\pm$ 0.10    & -1.91 $\pm$ 0.06   & -2.20 $\pm$ 0.18   & 1.61 $\pm$ 0.03                          & 1.621 $\pm$ 0.029                       \\
NGC 300                 & No-emission                                                                 & 9.33 $\pm$ 0.16     & 8.01 $\pm$ 0.18    & -2.62 $\pm$ 0.08   & -1.42 $\pm$ 0.08   & 1.43 $\pm$ 0.09                          & 2.157 $\pm$ 0.017                       \\
NGC 1407                & No-emission                                                                  & 10.023 $\pm$ 0.007  & ----               & -1.301 $\pm$ 0.006 & ----               & 2.509 $\pm$ 0.006                        & ----                                    \\
NGC 1549                & No-emission                                                                 & 10.013 $\pm$ 0.015  & 9.996 $\pm$ 0.014  & -1.326 $\pm$ 0.017 & -1.437 $\pm$ 0.017 & 2.3766 $\pm$ 0.0016                      & 2.3642 $\pm$ 0.0014                     \\
NGC 3923                & No-emission                                                                  & 9.849 $\pm$ 0.006   & 10.015 $\pm$ 0.015 & -1.308 $\pm$ 0.007 & -1.301 $\pm$ 0.005 & 2.4711 $\pm$ 0.0019                      & 2.4462 $\pm$ 0.0016                     \\
NGC 2835                & Unclassified                                                                 & 7.32 $\pm$ 0.28     & 7.86 $\pm$ 0.27    & -1.95 $\pm$ 0.24   & -1.74 $\pm$ 0.27   & 1.59 $\pm$ 0.03                          & 1.59 $\pm$ 0.03                         \\
NGC 4030                & Unclassified                                                                 & 9.76 $\pm$ 0.06     & 9.79 $\pm$ 0.05    & -1.47 $\pm$ 0.05   & -1.57 $\pm$ 0.04   & 1.932 $\pm$ 0.014                        & 1.982 $\pm$ 0.013                       \\
NGC 5247                & Unclassified                                                                 & 8.26 $\pm$ 0.17     & 8.22 $\pm$ 0.25    & -1.87 $\pm$ 0.21   & -2.29 $\pm$ 0.14   & 1.657 $\pm$ 0.027                        & 1.730 $\pm$ 0.021                       \\ \hline
\end{tabular}%
}
\end{table*}

\section{Third-degree polynomial fitting}\label{app_3d}

\begin{figure*}
\begin{center}

  \includegraphics[scale=0.22]{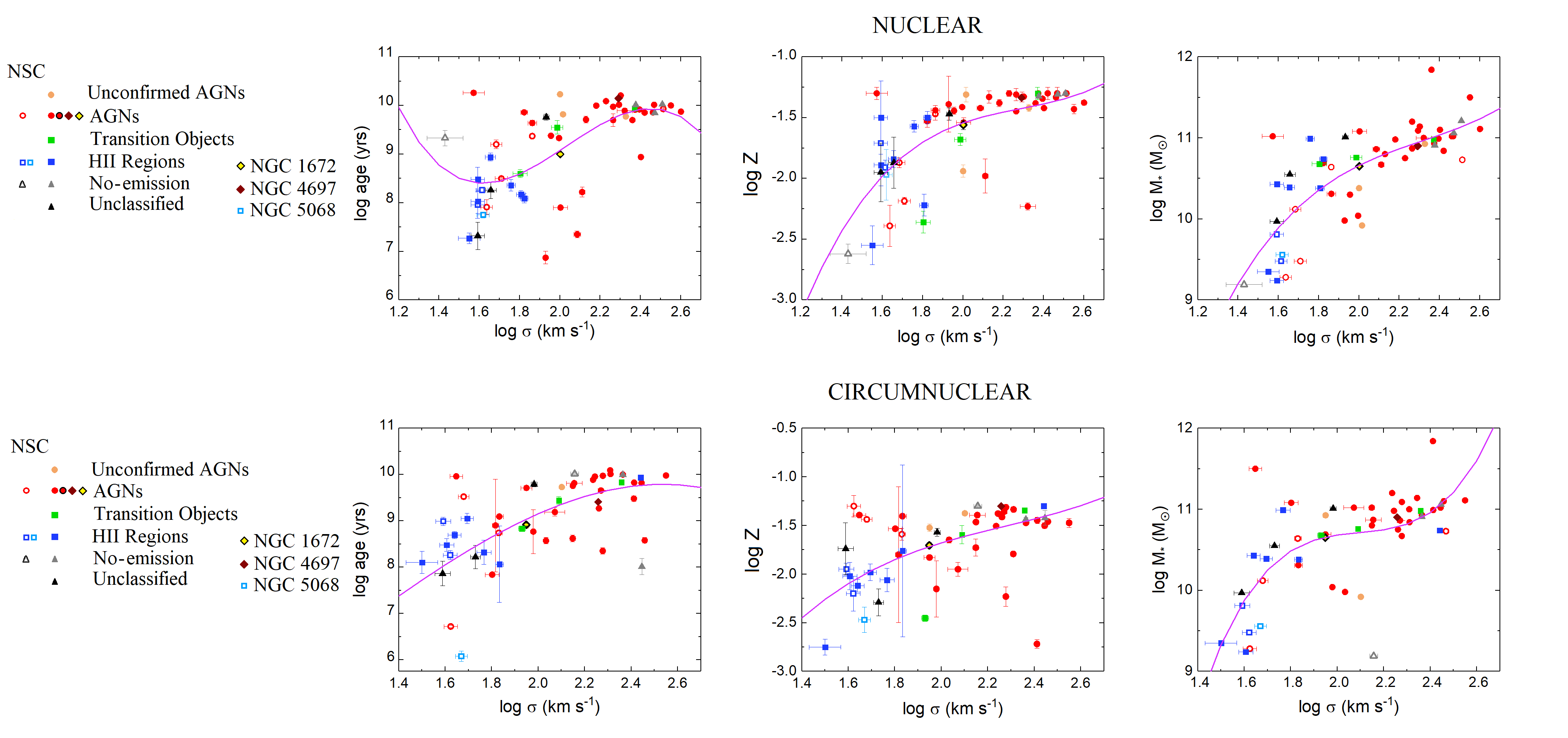}
  \caption{Third-degree polynomial fit (purple line) of the age, metallicity and mass versus the stellar velocity dispersion in the nuclear (first row of graphics) and in the circumnuclear (second row of graphics) spectra of the sample.} \label{plots_poly}
  
\end{center}
\end{figure*}

In order to evaluate more the correlations we observed in Figs. \ref{sigma_plots_nuclear} and \ref{sigma_plots_circumnuclear}, we applied third-degree polynomial fits to the first panels of those figures (since the linear fits are in the same first panels) shown in Fig. \ref{plots_poly}. The coefficients of each fit are shown in Table \ref{tablepoly}. The process is analogous to the linear fitting presented before.

Although the polynomial fits cover the distribution of the points better than the linear fits, it is difficult to determine which polynomial degree would be appropriate, as a higher-degree polynomial generally results in a better fit. In Fig. \ref{plots_poly}, we show third-degree polynomial fits to illustrate that such fitting is possible and delineates the correlation better than the linear fits. However, due to its complexity, we initially opted for the linear fit. In any case, the correlation is clear.

\begin{table*}[!ht]
\centering
\caption{Coefficients of the third-degree polynomial fit of the plots of Fig.\ref{plots_poly}.}
\label{tablepoly}
\resizebox{0.8\textwidth}{!}{%
\begin{tabular}{ccccccc}
\hline
\multicolumn{7}{c}{y = $A + Bx +Cx^2 +Dx^3$} \\ \hline
\multicolumn{4}{c}{Nuclear} & \multicolumn{3}{c}{Circumnuclear} \\ \hline
 & log(age) $\times$ log$\sigma$ & Z $\times$ log$\sigma$ & log (M$_*$) $\times$ log$\sigma$ &  log(age) $\times$ log$\sigma$ & Z $\times$ log$\sigma$  & log (M$_*$) $\times$ log$\sigma$ \\ \hline
A & (4.9 $\pm$ 2.8) $\times$ 10 & (-1.4 $\pm$ 1.1) $\times$ 10 & (-0.7 $\pm$ 1.6) $\times$ 10 & (0.3 $\pm$ 3.6) $\times$ 10 & (-1.1 $\pm$ 1.7) $\times$ 10 & (-4.3 $\pm$ 2.7) $\times$ 10 \\
B & (-6 $\pm$ 4) $\times$ 10 & (1.6 $\pm$ 1.6) $\times$ 10 & (2.2 $\pm$ 2.4) $\times$ 10 & (0 $\pm$ 5) $\times$ 10 & (1.1 $\pm$ 2.6) $\times$ 10 & (8 $\pm$ 4) $\times$ 10 \\
C & (3.3 $\pm$ 2.1) $\times$ 10 & -7 $\pm$ 8 & (-0.9 $\pm$ 1.2) $\times$ 10 & (0.2 $\pm$ 2.7) $\times$ 10 & (0.5 $\pm$ 1.3) $\times$ 10 & (-3.7 $\pm$ 2.0) $\times$ 10 \\
D & -5 $\pm$ 4 & 1.0 $\pm$ 1.3 & 1.3 $\pm$ 1.9 & -1 $\pm$ 4 & 0.7 $\pm$ 2.1 & 6 $\pm$ 3 \\ \hline
\end{tabular}%
}
\end{table*}

\end{appendix}


\end{document}